\documentclass[onecolumn,showpacs,nofootinbib,prd,amssymb]{revtex4-2}

\usepackage{latexsym}
\usepackage{graphicx,epsf, epsfig, amssymb, amsmath}

\usepackage[usenames,dvipsnames]{xcolor}
\usepackage{hyperref}

\usepackage{relsize}
\usepackage{todonotes}
\usepackage{textcomp}
\usepackage{float}  
\usepackage[normalem]{ulem}

\newcommand{\mnras}{Mon.\ Not.\ R.\ Astron.\ Soc.}
\newcommand{\physrep}{Phys.\ Rep.}

\newcommand{\araa}{Ann.\ Rev.\ Astron.\ Astrophys.}
\newcommand{\apss}{Astrophys.\ Space\ Sci.}
\newcommand{\nphysa}{Nucl.~Phys.~A}          

\DeclareMathAlphabet\mathbfcal{OMS}{cmsy}{b}{n}

\newcommand{\pd}[2]{\frac{\partial #1}{\partial #2}} 

\newcommand{\nablaperp}{\mathbin{^{\mathsmaller{\perp}}\mkern-3mu \nabla}}

\newcommand{\Fperp}{\mathbin{^{\mathsmaller{\perp}}\mkern-3mu F}}

\newcommand{\Gperp}{\mathbin{^{\mathsmaller{\perp}}\mkern-2mu G}}

\newcommand{\Vperp}{\mathbin{^{\mathsmaller{\perp}}\mkern-1mu \mathcal{V}}}
\newcommand{\Vpar}{\mathbin{^{\mathsmaller{\parallel}}\mkern-1mu \mathcal{V}}}

\newcommand{\Wperp}{\mathbin{^{\mathsmaller{\perp}}\mkern-1mu \mathcal{W}}}
\newcommand{\Wpar}{\mathbin{^{\mathsmaller{\parallel}}\mkern-1mu \mathcal{W}}}

\newcommand{\Xperp}{\mathbin{^{\mathsmaller{\perp}}\mkern-6mu \mathcal{X}}}
\newcommand{\Xpar}{\mathbin{^{\mathsmaller{\parallel}}\mkern-6mu \mathcal{X}}}

\begin{document}

\title{Dissipative superfluid relativistic magnetohydrodynamics of a multicomponent fluid:
	the combined effect of particle diffusion and vortices}
\author{V.~A.~Dommes}
\email[e-mail: ]{vasdommes@gmail.com}
\author{M.~E.~Gusakov}
\affiliation{Ioffe Institute,
Politekhnicheskaya 26, 194021 St.~Petersburg, Russia
}

\begin{abstract} 
We formulate dissipative magnetohydrodynamic equations for
finite-temperature
superfluid and superconducting
charged relativistic mixtures,
taking into account 
the effects of
particle diffusion
and 
possible presence 
of Feynman-Onsager and/or Abrikosov vortices in the system.
The equations depend on a number of phenomenological transport coefficients, which describe, in particular, 
relative 
motions of different particle species
and their interaction with vortices. 
We demonstrate how to relate
these transport coefficients 
to the mutual friction parameters and momentum transfer rates
arising in the
microscopic theory.
The resulting equations can be
used to study, in a unified and coherent way, a very wide range of phenomena associated with dynamical processes in neutron stars, e.g., 
the magnetothermal evolution, stellar oscillations and damping, as well as development and suppression of various hydrodynamic instabilities in neutron stars.
\end{abstract}

\date{\today}

%
%
%
%
%
%
%
%
%
%



\maketitle

\section{Introduction}
\label{sec:intro}

Consider a dense mixture composed of several particle species, some of which may be charged.
Assume also that some components of the mixture are in a superfluid and/or superconducting state at finite temperature.
In what follows, we are interested in describing the behavior of such system in the {\it hydrodynamic} regime,
i.e., assuming that the typical particle mean-free path and collision time
are much smaller than, respectively, the typical lengthscale and timescale of the evolution of the system.

Assume further that:
(i) the mixture is relativistic, 
and can be 
in a strong gravitational field;
(ii) the mixture is magnetized and rotating, so that there are Feynman-Onsager and Abrikosov vortices in the system
	(below we assume that the charged superconducting particles form a type-II superconductor);
(iii) normal (nonsuperfluid and nonsuperconducting) particles of different species
	do {\it not} move with exactly the same velocities,
	in other words, we allow for the {\it diffusion} of normal particles with respect to 
	each other.
Then, the question is, what are the equations describing dynamics in such a system?

Before answering this question (which is the subject of the present study)
let us explain why it is important for us to formulate such equations.
The reason is that  mixtures with the properties 
just described
can be found
in the inner layers (cores) of neutron stars (NSs).
An NS core consists, in the simplest case,
of neutrons ($n$), protons ($p$), and electrons ($e$) with an admixture of muons ($\mu$).
This matter is extremely compact and degenerate --
its density is several times greater than the density of matter in atomic nuclei,
$\rho_0=2.8 \times 10^{14}$~g~cm$^{-3}$.
Magnetic fields in NSs may reach enormous values
$\gtrsim 10^{15}$~G 
\cite{kb17,bl16},
while the
gravitational field is so strong that the NS radius ($\sim 10$~km)
is only a few times larger than the Schwarzschild radius \cite{hpy07}.
Furthermore, according to microscopic calculations \cite{sc19,ding16,gps14,ls01}, as well as
observations of 
cooling, glitching, and rapidly rotating NSs
\cite{yp04,plps13,kgd20,sohh21,hm15},
baryons (in particular, neutrons, and protons)
in NS interiors are expected to become superfluid/superconducting
at temperatures $T \lesssim 10^8 - 10^{10}~{\rm K}$.
This means that, if an NS is rotating and magnetized,
the topological defects -- neutron (Feynman-Onsager) vortices
and proton (Abrikosov) flux tubes -- may be present (and co-exist) in the system \cite{sauls89,hs18}.%
\footnote{Here we assume that protons form a type-II superconductor, which is likely true
for the outer part of the NS core, but, probably, not
the case for the inner part \cite{sedrakian05,wgn20}.
}
%
The equations presented in this paper are designed precisely to describe
various dynamical 
phenomena
in NSs, such as NS oscillations, cooling, and magnetic field evolution.

Our paper is, of course, not the first one in a series of works that have studied the dynamics of such systems.
The smooth-averaged nonrelativistic hydrodynamics describing superfluid liquid helium-II with vortices
was formulated by Hall and Vinen \cite{hv56,hall60} and, independently, by Bekarevich and Khalatnikov \cite{bk61}.
It has been extended in subsequent studies (e.g., \cite{vs81,ml91,ss95,cl98,gas11,gusakov16,gd16,awv16,ga20,rw20,andersson21}) to account for charged mixtures and relativistic effects.
Recently, Ref.~\cite{gd16} (hereafter GD16) derived 
the relativistic MHD, which describes superfluid/superconducting mixtures at finite temperatures,
and allows for the presence of
Feynman-Onsager and Abrikosov vortices, as well as the electromagnetic field.
It focuses mainly on the nondissipative equations,
and ignores particle diffusion, viscosity, and other dissipative effects (except for the mutual friction dissipation, which
is taken into account).
This work was further extended by Rau and Wasserman \cite{rw20}
who obtained an equivalent formulation of relativistic MHD
starting from the Carter's variational principle \cite{carter91}, and also included heat conduction and viscosity
into the 
corresponding
equations.

All these works ignore particle diffusion, i.e., relative motions of different particle species
(or Bogoliubov thermal excitations, if superfluid/superconducting species are considered)
with respect to each other.
This is an unfortunate omission, since it is well-known
that diffusion plays a crucial role
in the secular evolution
of the magnetic field in nonsuperfluid and nonsuperconducting NSs \cite{gr92,su95,crv17,papm17,gko17,og18,crv20}
and, moreover,
can be very efficient \cite{kg20,kgk21} in
damping of NS oscillations and suppressing various instabilities in
their interiors.
As shown recently \cite{gko20},
diffusion also has a major effect on
the evolution of the magnetic field in {\it superconducting}
NSs.
The reason is easy to understand.
If protons form a type-II superconductor, the magnetic field in the NS core
is locked to quantized proton flux tubes
and its evolution is determined by the flux tube motion.
To study this motion, one has to calculate the balance of forces acting on vortices, which (except for the buoyancy and tension forces \cite{hall60,bk61,mt85,dg17})
depend on the relative velocities between vortices and different particle species that scatter on it.
Because interaction (in particular, friction)
of particles with vortices is very strong due to the huge amount of vortices in the system \cite{gusakov19,gko20},
even small mismatch in the velocities of different particle species
significantly affects the force balance on vortices, and hence the magnetic field evolution.

Up until now the MHD equations, describing relativistic charged mixtures,
and systematically incorporating the diffusion effects
have been studied
in the very limited number of works and only neglecting the superconductivity and superfluidity effects.
In particular, the most advanced MHD versions, suitable for NS modeling,
were formulated
in the series of papers by Andersson et al. \cite{andersson12,ach17,ahdc17,adhc17}, and in Ref.~\cite{dgs20} (hereafter DGS20).
In the present work we fill this gap
by combining the results of GD16 and DGS20,
with the aim
to formulate the ready-to-use dissipative relativistic MHD for superfluid/superconducting mixtures,
accounting for both
vortices
and diffusion effects.
We follow the same 
approach \cite{bk61,ll87} as in those papers.
Namely, we build a first-order dissipative hydrodynamics, starting from the conservation laws
and then deriving the general form of dissipative terms,
which are (i) linear in thermodynamic fluxes,
(ii) ensure non-negative entropy production rate,
and (iii) satisfy the Onsager relations.
The first-order MHD formulated in this paper is strictly valid in the hydrodynamic regime, i.e.,
as long as the typical lengthscale and timescale in the problem 
are much larger than the particle mean-free path and collision time, respectively.
Although we did not test our MHD, 
it 
has been argued in the literature
(e.g., \cite{hl83, hl85}) 
that a generic
first-order theory may have theoretical issues
with acausality and stability due to \underline{unphysical} 
high-frequency/short-wavelength modes,
which lie outside 
the applicability domain of the hydrodynamic regime.
One way to overcome these issues 
is to use more
complicated formulations, 
such as the first-order theories with a specially chosen reference frame \cite{kovtun19}, 
second-order theories \cite{israel76,is79,lsmr86}, 
or hydrodynamics based on the Carter's variational principle \cite{carter91,andersson12,adhc17}.%
\footnote{
    Note that in the hydrodynamic regime the higher-order corrections are typically small.
    This is clearly illustrated in section VIII of DGS20,
    where it is shown that such corrections to the standard (acausal) heat equation
    can be safely ignored.
}
The other (less elegant, but
more pragmatic)
option, 
which applies to those who work in the deep hydrodynamic regime, 
is simply to discard the unphysical modes in the solution, or filter them out,
when it comes to numerical implementation.
Moreover, for many practical applications, 
where the MHD formulated in this work can be used
(e.g., modeling the NS magnetothermal evolution or oscillations
and related physical instabilities),
the macroscopic particle velocities appear to be nonrelativistic.
Then
the relativistic equations
(see, e.g., section V of DGS20 and Appendix \ref{sec:mhd-full-nonrel})
have a similar structure to 
the nonrelativistic ones;
the main difference is the relativistic equation of state
and, if one allows for the effects of general relativity, the metric coefficients.
In this case
additional degrees of freedom 
(which arise in the relativistic treatment
and do not have Newtonian counterparts)
are absent, and thus the
hydrodynamics
remains stable \cite{gah20}.
Bearing in mind the above 
comments,
we leave detailed 
discussion
of theoretical acausality and instability issues 
beyond the scope of the present
work.

The paper is organized as follows.
In Sec.~\ref{sec:gen-eqns} we
formulate general hydrodynamics equations for charged
superfluid/superconducting relativistic mixtures in the presence of vortices
and the electromagnetic field, accounting for a number of dissipative effects:
mutual friction, diffusion, viscosity, and chemical reactions.
In Sec.~\ref{sec:entropy}, we derive the entropy generation equation
and in Sec.~\ref{sec:f-dj} we use 
it together with the
Onsager relations to derive the general form of
dissipative corrections for particle currents, as well as mutual friction forces acting on vortices.
In Sec.~\ref{sec:special-cases} we apply these general formulas to a number of interesting limiting cases, which are suitable for NS applications.
Sec.~\ref{sec:mhd-full} provides a full set of hydrodynamic equations
in the ``MHD approximation'' 
adopted in GD16,
which is applicable
for typical NS conditions and allows one to study a long-term 
magnetothermal evolution
in superconducting NSs.
Finally, we sum up in Sec.~\ref{sec:conclusion}.
The paper also contains two appendices.
Appendix~\ref{sec:mhd-full-nonrel} presents a nonrelativistic limit of MHD equations
from Sec.~\ref{sec:mhd-full}.
In Appendix~\ref{sec:mf} we show how to express the phenomenological transport coefficients appearing in our equations through
the mutual friction
parameters
and momentum transfer rates calculated from the microscopic theory.

Unless otherwise stated, in what follows 
the speed of light $c$
and the Boltzmann constant $k_{\rm B}$ are set to unity,
$c=k_{\rm B}=1$.

\section{General equations}
\label{sec:gen-eqns}

In this section, we present dissipative equations,
describing dynamics of charged finite-temperature superfluid relativistic mixtures
in the presence of vortices in the {\it hydrodynamic} regime (see Introduction).
For definiteness, and bearing in mind NS applications,
we consider a mixture composed of superfluid neutrons, superconducting protons,
normal electrons, and normal muons.%
\footnote{We do not assume that all neutrons and protons are necessarily
in the Cooper-pair condensate. In other words, we allow for the possible presence
of normal neutron and proton component in the mixture.}
%
Both neutron (Feynman-Onsager) vortices and proton (Abrikosov) flux tubes can be present in the system.
Generalization of these equations to more complex compositions (e.g., including hyperons) is straightforward.

The dynamical equations proposed
here are
very similar to those formulated in
GD16
assuming
type-II proton superconductivity,
but contain a number of extra terms:
(i) the four-force $G^\nu$ in the right-hand side of Eq.~\eqref{eq:dTmunu=0};
(ii) the particle production rate $\Delta\Gamma_i$ in the right-hand side of Eq.~\eqref{eq:continuity};
(iii) the dissipative correction $\Delta j_{(i)}^{\mu}$ to the particle current density \eqref{eq:jmu}; 
(iv) the dissipative correction $\Delta \tau^{\mu\nu}$ to the energy-momentum tensor \eqref{eq:Tmunu};
and (v) the superfluid dissipative correction $\varkappa_i$ to the chemical potential $\mu_i$
in the definitions \eqref{eq:w} and \eqref{eq:Vmunu}.
Note that the first four corrections are included in the nonsuperfluid dissipative MHD of DGS20,
but for superfluid/superconducting mixtures their actual form may differ.

\vspace{0.2 cm}
\noindent
%
{\bf Continuity equations}

The four-current density $j_{(i)}^\mu$
of particle species $i$
satisfies
the continuity equation
\begin{gather}
\label{eq:continuity}
	\partial_\mu j_{(i)}^{\mu} = \Delta\Gamma_i
,
\end{gather}
where $\partial_\mu \equiv \partial/\partial x^{\mu}$ is the four-gradient,
and $\Delta\Gamma_i$ is the corresponding production rate (source of particles $i$).
Here and below, unless otherwise stated, Latin indices $i,k,\ldots$ refer to particle species
(neutrons $n$, protons $p$, electrons $e$, and muons $\mu$),
whereas Greek letters $\mu,\nu\ldots = 0,1,2,3$ denote the space-time indices,
and summation over repeated indices is assumed.

In the simplest case
of nonsuperfluid matter in the absence of diffusion,
the particle current density is $j_{(i)}^{\mu} = {n_i u^{\mu}}$,
where $u^{\mu}$ is the (common for all particle species)
normal four-velocity, normalized by the condition
\begin{gather}
\label{eq:uu=-1}
	u_{\mu} u^{\mu} = -1
,
\end{gather}
and $n_i$ is the particle number density measured in the comoving frame
$u^\mu = (1,0,0,0)$, such that
\begin{gather}
\label{eq:uj=-n}
	u_{\mu} j_{(i)}^{\mu} = -n_i
.
\end{gather}

When accounting for superfluidity and diffusive currents,
$j_{(i)}^{\mu}$ can generally be presented as a sum of three terms:
\begin{gather}
\label{eq:jmu}
	j_{(i)}^{\mu} = {n_i u^{\mu}} + Y_{ik} w_{(k)}^\mu + \Delta j_{(i)}^{\mu}
,
\end{gather}
where the four-vector $w_{(k)}^\mu$
describes the superfluid degrees of freedom \cite{ga06}
and satisfies the condition \cite{ga06, gusakov07, gusakov16}
\begin{gather}
\label{eq:uw=0}
	u_\mu w_{(i)}^\mu = 0.
\end{gather}
This vector is related to the wave-function phase $\Phi_i$ of the Cooper condensate
by the formula
\begin{gather}
\label{eq:w}
	w_{(i)}^\mu = \partial^\mu \phi_i - \left( \mu_i + \varkappa_i \right) u^\mu - e_i A^\mu
,
\end{gather}
where $\partial^\mu \phi_i = (\hbar/2) \partial^\mu \Phi_i$ \cite{ga06},
$\hbar$ is the Planck constant,
$\mu_i$ is the relativistic chemical potential for particle species $i$,
$A^\mu$ is the electromagnetic potential,
and $\varkappa_i$ is the viscous dissipative correction
to the chemical potential \cite{gusakov07,gusakov16}.

Further, $Y_{ik}$ in Eq.\ \eqref{eq:jmu}
is the symmetric entrainment matrix \cite{ga06,gkh09a,gkh09b,gusakov10, ghk14},
which is a relativistic analogue of the nonrelativistic superfluid mass-density matrix \cite{ab76,gh05,leinson18,ac21};
and $\Delta j_{(i)}^{\mu}$ is the dissipative correction
due to nonsuperfluid diffusive currents (see DGS20 for a similar definition of $\Delta j_{(i)}^{\mu}$ in normal matter).

Throughout the paper, all the thermodynamic quantities are defined (measured)
in the comoving frame.
This means that the relation \eqref{eq:uj=-n}  holds
also
in the general case (when dissipation effects are allowed for),
which imposes
an additional constraint on
$\Delta j_{(i)}^{\mu}$,
\begin{gather}
\label{eq:udj=0}
	u_{\mu} \Delta j_{(i)}^{\mu} = 0
.
\end{gather}
%

\vspace{0.2 cm}
\noindent
%
{\bf Energy-momentum conservation }

The relativistic energy-momentum conservation law
takes the form
\begin{gather}
\label{eq:dTmunu=0}
	\partial_{\mu} T^{\mu\nu} = G^\nu
,
\end{gather}
where $G^\nu$ is the radiation four-force density,
which describes exchange of energy and momentum between matter and radiation%
%
\footnote{
	For isotropic emission $G^\nu = - Q u^\nu$, where $Q$
	is the total emissivity
	(e.g., it can be the neutrino emissivity
	due to beta-processes in the NS core).},
%
and the energy-momentum tensor $T^{\mu\nu}$ is given by
\begin{gather}
\label{eq:Tmunu}
	T^{\mu\nu} =
		{(P+\varepsilon) u^{\mu} u^\nu}
		+ {P g^{\mu\nu}} 
		+ Y_{ik} \left(
			w_{(i)}^\mu w_{(k)}^\nu
			+ \mu_i w_{(k)}^\mu u^\nu
			+ \mu_k w_{(i)}^\nu u^\mu
			\right)
		+ \Delta T^{\mu\nu}_{({\rm EM+vortex})} + \Delta \tau^{\mu\nu}
,\end{gather}
where $P$ is the pressure defined by Eq.~\eqref{eq:pres} below,
$\varepsilon$ is the energy density,
and $g_{\mu\nu}={\rm diag}(-1,1,1,1)$ is the space-time metric.%
\footnote{
	In this paper, we assume that the metric is flat.
	Our results can easily be generalized to an arbitrary metric,
	provided that all relevant length scales
	are much smaller than the characteristic gravitational lengthscale.
	In this case, one has to replace
	all ordinary derivatives with their covariant counterparts and, in addition,
	replace the Levi-Civita tensor
	$\epsilon^{\mu\nu\lambda\sigma}$
	with $\eta^{\mu\nu\lambda\sigma} \equiv \left(- {\rm det}~g_{\alpha\beta} \right)^{-1/2} ~ \epsilon^{\mu\nu\lambda\sigma}$.
}
The energy-momentum tensor \eqref{eq:Tmunu} is a sum of
the energy-momentum tensor
of a vortex-free uncharged superfluid hydrodynamics (the first three terms)
plus electromagnetic and vortex contributions $\Delta T^{\mu\nu}_{({\rm EM+vortex})}$ given by Eq.~\eqref{eq:dT-EM-vortex} below,
and dissipative correction $\Delta \tau^{\mu\nu}$.
Note that all these terms except for the last one are the same as in GD16.

In the comoving frame the energy density is given by the component $T^{00}$ of the energy-momentum tensor,
$T^{00}=\varepsilon$,
which implies
\begin{gather}
\label{eq:uuT=e}
	u_{\mu}u_{\nu} T^{\mu\nu} = \varepsilon
.
\end{gather}
This relation, in view of the expressions \eqref{eq:Tmunu}, \eqref{eq:uw=0} \eqref{eq:dT-EM-vortex}--\eqref{eq:TM}, \eqref{eq:TVE}, and \eqref{eq:TVM},
imposes the following constraint
on the dissipative correction $\Delta \tau^{\mu\nu}$,
\begin{gather}
\label{eq:uudtau=0}
	u_\mu u_{\nu} \Delta \tau^{\mu\nu} = 0
.
\end{gather}
Note, however, that the four-velocity $u^\mu$ itself is not uniquely defined in the system with dissipation
(see, e.g., a thorough discussion of a similar issue in Ref. \cite{ll87} and in DGS20).
We specify $u^\mu$ by requiring the total momentum
of the normal fluid component to be zero in the comoving frame.
This leads to an
additional condition for $\Delta \tau^{\mu\nu}$,
\begin{gather}
\label{eq:udtau=0}
	u_{\nu} \Delta \tau^{\mu\nu} = 0.
\end{gather}
The condition (\ref{eq:udtau=0}) coincides with 
the similar condition
defining the
so-called Landau-Lifshitz
(or transverse) frame
of nonsuperfluid relativistic hydrodynamics \cite{ll87}.

\vspace{0.2 cm}
\noindent
%
{\bf Maxwell equations}

Electromagnetic field is described by the Maxwell equations in the medium,
\begin{gather}
\label{eq:divD}
	{\rm div} {\pmb D} = 4 \pi \rho_{\rm free}
,\\
\label{eq:curlE}
	{\rm curl} {\pmb E} 
	= - \pd{{\pmb B}}{t}
,\\
\label{eq:divB}
	{\rm div} {\pmb B} = 0
,\\
\label{eq:curlH}
	{\rm curl} {\pmb H} = 4 \pi {\pmb J}_{\rm free} + \pd{{\pmb D}}{t}
,
\end{gather}
where
${\pmb E}$ is the electric field,
${\pmb B}$ is the magnetic induction,
${\pmb D}$ is the electric displacement,
${\pmb H}$ is the magnetic field,
$\rho_{\rm free}$ is the free charge density,
and ${\pmb J}_{\rm free}$ is the current density
of free charges.
Note that, generally,
${\pmb D} \neq {\pmb E}$ and ${\pmb H} \neq {\pmb B}$,
since there are bound charges and bound currents in the system,
associated with superfluid/superconducting vortices
and their motion (for details see GD16);
in the absence of vortices
(and neglecting very weak magnetization and polarizability of NS matter \cite{bpl00})
${\pmb D} = {\pmb E}$ and ${\pmb H} = {\pmb B}$.

The explicitly covariant form of Maxwell equations \eqref{eq:divD}--\eqref{eq:curlH}
is \cite{ll60,toptygin15}
\begin{gather}
\label{eq:maxwell-1}
	\partial_{\mu} F_{\nu\lambda}
	+ \partial_{\nu} F_{\lambda\mu}
	+ \partial_{\lambda} F_{\mu\nu}
	= 0
,\\
\label{eq:maxwell-2}
	\partial_\nu G^{\mu\nu}
	= 4\pi J^{\mu}_{\rm (free)}
,
\end{gather}
where the antisymmetric electromagnetic tensors $F^{\mu\nu} \equiv \partial^\mu A^\nu - \partial^\nu A^\mu$
and $G^{\mu\nu}$
are composed of components of the vectors ${\pmb E}$, ${\pmb B}$, ${\pmb D}$, and ${\pmb H}$,
\begin{gather}
F^{\mu\nu}
= 
\left(
\begin{array}{cccc}
	0 & E_1 & E_2 & E_3
	\\
	-E_1 & 0 & B_3 & -B_2
	\\
	-E_2 & -B_3 & 0 & B_1
	\\
	-E_3 & B_2 & -B_1 & 0
\end{array}
\right)
,\\
G^{\mu\nu}
= 
\left(
\begin{array}{cccc}
	0 & D_1 & D_2 & D_3
	\\
	-D_1 & 0 & H_3 & -H_2
	\\
	-D_2 & -H_3 & 0 & H_1
	\\
	-D_3 & H_2 & -H_1 & 0
\end{array}
\right)
,
\end{gather}
and $J^{\mu}_{\rm (free)} = \left( \rho_{\rm free}, {\pmb J}_{\rm free} \right)$
is the four-current density of free charges,
\begin{gather}
\label{eq:Jfree}
	J^{\mu}_{\rm (free)}
	\equiv e_i j_{(i)}^\mu
	= e_i n_i u^\mu
		+ e_i Y_{ik} w_{(k)}^\mu
		+ e_i \Delta j_{(i)}^\mu
,
\end{gather}
where $e_i$ is the electric charge for particle species $i$.

\vspace{0.2 cm}
\noindent
%
{\bf Vorticity tensor}

Following GD16, we introduce the vorticity tensor
\begin{gather}
\label{eq:Vmunu}
	\mathcal{V}^{\mu\nu}_{(i)}
	\equiv
		 \partial^\mu\left[ w^\nu_{(i)}+ (\mu_i + \varkappa_i) u^{\nu} + e_i A^{\nu}\right]
		-\partial^\nu\left[ w^\mu_{(i)}+(\mu_i + \varkappa_i) u^{\mu}+ e_i A^{\mu}\right]
,
\end{gather}
which is a relativistic generalization of the three-vector $m_i {\rm curl}{\pmb V}_{{\rm s}i} + (e_i / c) {\pmb B}$ (see Appendix~\ref{sec:mhd-full-nonrel}).
In a system without topological defects (i.e., vortices),
the superfluid phase $\Phi_i$ is a smooth function of coordinates
satisfying the condition
$\partial^\mu \partial^\nu \Phi_i -\partial^\nu \partial^\mu \Phi_i = 0$,
which, in view of Eq.~\eqref{eq:w}, translates into
\begin{gather}
\label{eq:Vmunu=0}
	\mathcal{V}^{\mu\nu}_{(i)} = 0
.
\end{gather}

However, in the presence of vortices, the condition
$\partial^\mu \partial^\nu \Phi_i -\partial^\nu \partial^\mu \Phi_i = 0$
is violated at the vortex lines.
Consequently, the (smooth-averaged) vorticity tensor $\mathcal{V}^{\mu\nu}_{(i)}$ differs from zero.
One can demonstrate that this tensor $\mathcal{V}^{\mu\nu}_{(i)}$
is related to the number of vortices ${\mathcal N}_{{\rm V}i}$ piercing the closed contour
by the relation \cite{gusakov16}\footnote{
This relation is satisfied
for Fermi superfluids (e.g., neutrons or protons);
for Bose superfluids there should be $2\pi \hbar {\mathcal N}_{{\rm V}i}$ in the right-hand side of the equation.
Note that the factor $1/2$ was inadvertently omitted in the corresponding equation~(42) in Ref.~\cite{gusakov16}.
}
\begin{gather}
\label{eq:Nvi}
	\frac{1}{2} \, \int d f^{\mu \nu} \, 
	\mathcal{V}_{(i)\mu \nu}
	= \pi \hbar {\mathcal N}_{{\rm V}i}
.
\end{gather}
Eq.~\eqref{eq:Vmunu=0} then should be replaced by a more general superfluid equation \eqref{eq:sfl-eqn} introduced in Sec.~\ref{sec:f-dj} below.

\vspace{0.2 cm}
\noindent
%
{\bf Thermodynamic relations}

The dynamic equations listed above should be supplemented by the second law of thermodynamics,
\begin{equation}
\label{eq:2ndlaw}
	d \varepsilon = \mu_i \, dn_i 
		+ T \, dS 
		+ \frac{Y_{ik}}{2} \, d \left( w_{(i)}^\alpha w_{(k)\alpha}  \right)
		+ d \varepsilon_{\rm add}
,
\end{equation}
where 
$T$ is the temperature,
$S$ is the entropy per unit volume,
and the electromagnetic/vortex contribution to the energy density 
$d \varepsilon_{\rm add}$ reads [see equation (79) in GD16]
\begin{gather}
\label{eq:de-EM-vortex}
	d \varepsilon_{\rm add}
	= 
	\frac{1}{4\pi} \, E_{\mu} d D^{\mu} + \frac{1}{4\pi} \, H_{\mu} d B^{\mu}
	+ \mathcal{V}^{\mu}_{({\rm E}i)} d \mathcal{W}_{({\rm E}i)\mu}
	+ \mathcal{W}_{({\rm M}i)\mu} d \mathcal{V}^{\mu}_{({\rm M}i)}
.
\end{gather}
Here we introduced the auxiliary vortex-related vectors
$\mathcal{W}_{({\rm E}i)}^{\mu}$
and 
$\mathcal{W}_{({\rm M}i)}^{\mu}$,
in full analogy with the electromagnetic vectors
$D^{\mu}$
and
$H^{\mu}$,
respectively.
Eq. \eqref{eq:de-EM-vortex} should be considered as a {\it definition} of the vectors
$D^{\mu}$, $H^{\mu}$, $\mathcal{W}_{({\rm E}i)}^{\mu}$,
and 
$\mathcal{W}_{({\rm M}i)}^{\mu}$
[or, equivalently, the tensors
$G^{\mu\nu}$
and
$\mathcal{W}_{(i)}^{\mu \nu}$,
see the identities
\eqref{eq:Emu}--\eqref{eq:Wmagn}
below].
When a microscopic model for the system energy density is specified
(see, e.g., Appendix G in GD16
and Sec.~\ref{sec:mhd-limit}),
one can express these vectors through the vectors
$E^{\mu}$, $B^{\mu}$, $\mathcal{V}_{({\rm E}i)\mu}$,
and 
$\mathcal{V}_{({\rm M}i)\mu}$
(or, equivalently, through the tensors
$F^{\mu\nu}$
and
$\mathcal{V}_{(i)}^{\mu \nu}$).
The four-vectors entering Eq.~\eqref{eq:de-EM-vortex}
are related to the corresponding tensors as
\begin{gather}
\label{eq:Emu}
	E^\mu \equiv u_\nu F^{\mu\nu}
,\\
\label{eq:Dmu}
	D^\mu \equiv u_\nu G^{\mu\nu}
,\\
\label{eq:Bmu}
	B^\mu \equiv \frac{1}{2} \epsilon^{\mu\nu\alpha\beta} u_\nu F_{\alpha\beta}
,\\
\label{eq:Hmu}
	H^\mu \equiv \frac{1}{2} \epsilon^{\mu\nu\alpha\beta} u_\nu G_{\alpha\beta}
,\\
\label{eq:Velectr}
	\mathcal{V}^{\mu}_{({\rm E}i)}
	\equiv  u_\nu \mathcal{V}^{\mu\nu}_{(i)}
,\\
\label{eq:Vmagn}
	\mathcal{V}^{\mu}_{({\rm M}i)}
	\equiv
	\frac{1}{2} \, \epsilon^{\mu \nu \alpha \beta} \, u_{\nu} \, \mathcal{V}_{(i) \alpha \beta}
,\\
\label{eq:Welectr}
	\mathcal{W}^{\mu}_{({\rm E}i)} \equiv  
	u_\nu \mathcal{W}^{\mu\nu}_{(i)}
,\\
\label{eq:Wmagn}
	\mathcal{W}^{\mu}_{({\rm M}i)} 
	\equiv
	\frac{1}{2} \, \epsilon^{\mu \nu \alpha \beta} \, u_{\nu} \, \mathcal{W}_{(i) \alpha \beta}
,
\end{gather}
where 
the Levi-Civita tensor $\epsilon^{\mu\nu\alpha\beta}$ is defined such that $\epsilon^{0123}=1$.
In the comoving frame, $u^\mu = (1,0,0,0)$,
the four-vectors $E^\mu$, $D^\mu$, $B^\mu$ and $H^\mu$ are
related to, respectively,
the ordinary three-vectors ${\pmb E}$, ${\pmb D}$, ${\pmb B}$, and ${\pmb H}$ as:
$E^\mu = (0, {\pmb E})$,
$D^\mu = (0, {\pmb D})$,
$B^\mu = (0, {\pmb B})$,
$H^\mu = (0, {\pmb H})$.

The total pressure $P$
is defined (see, e.g., GD16) as
a partial derivative of the full system energy 
$\varepsilon V$ with respect to the volume $V$ at constant
total number of particles $n_i V$, total entropy $SV$, 
as well as at fixed quantities
$w_{(i)}^\alpha w_{(k)\alpha}$,
$D^\mu$, $B^\mu$,
$\mathcal{W}^{\mu}_{({\rm E}i)}$, and 
$\mathcal{W}^{\mu}_{({\rm M}i)}$:
\begin{gather}
\label{eq:pres}
	P
	\equiv -\frac{\partial \left(\varepsilon V \right)}{\partial V}
	= -\varepsilon +\mu_i n_i + TS
,
\end{gather}
Using Eqs.\ \eqref{eq:2ndlaw}, \eqref{eq:de-EM-vortex}, and \eqref{eq:pres},
one arrives at the following Gibbs-Duhem relation,
\begin{equation}
\label{eq:dP}
	dP = n_i  \, d\mu_i + S \, dT
	- \frac{Y_{ik}}{2} \, d \left( w_{(i)}^\alpha w_{(k)\alpha}  \right)
	- \frac{1}{4\pi} E_\alpha dD^\alpha 
	- \frac{1}{4\pi} H_\alpha dB^\alpha
	- \mathcal{V}^{\mu}_{({\rm E}i)} d \mathcal{W}_{({\rm E}i)\mu}
	- \mathcal{W}_{({\rm M}i)\mu} d \mathcal{V}^{\mu}_{({\rm M}i)}
.
\end{equation}

\vspace{0.2 cm}
\noindent
%
{\bf Electromagnetic and vortex contribution to $T^{\mu\nu}$}

The electromagnetic and vortex contribution to $T^{\mu\nu}$,
represented
by the term $\Delta T^{\mu\nu}_{({\rm EM + vortex})}$ in Eq.~\eqref{eq:Tmunu},
has been derived in GD16, and takes the form
\begin{gather}
\label{eq:dT-EM-vortex}
	\Delta T^{\mu\nu}_{({\rm EM + vortex})}
	= \mathcal{T}^{\mu\nu}_{({\rm E})}
	+ \mathcal{T}^{\mu\nu}_{({\rm M})}
	+ \mathcal{T}^{\mu\nu}_{({\rm VE})}
	+ \mathcal{T}^{\mu\nu}_{({\rm VM})}
,
\end{gather}
where the electromagnetic contributions $\mathcal{T}^{\mu\nu}_{({\rm E})}$ and $\mathcal{T}^{\mu\nu}_{({\rm M})}$
are given by [see equations (66) and (67) in GD16]
\begin{eqnarray}
\label{eq:TE}
\mathcal{T}^{\mu\nu}_{({\rm E})} &=& \frac{1}{4\pi} \, 
\left(
\perp^{\mu\nu} D^{\alpha}E_{\alpha}-D^{\mu}E^{\nu}
\right)
,\\
\label{eq:TM}
\mathcal{T}^{\mu\nu}_{({\rm M})} &=& 
\frac{1}{4\pi} 
\left(
\Gperp^{\mu\alpha}\Fperp^{\nu}_{\,\,\,\alpha} + u^{\nu} \Gperp^{\mu\alpha}E_{\alpha}
+u^{\mu}\Gperp^{\nu\alpha}E_{\alpha}
\right)
.
\end{eqnarray}
Here and hereafter $\perp^{\mu\nu} \equiv g^{\mu\nu} + u^\mu u^\nu$,
and the notation $\Xpar^{\mu\nu}$ and $\Xperp^{\mu\nu}$
is introduced
for arbitrary antisymmetric tensor $\mathcal{X}^{\mu\nu}$:
\begin{gather}
\label{eq:Apar}
	\Xpar^{\mu\nu}
	= -u^\nu \mathcal{X}_{({\rm E})}^\mu +u^\mu \mathcal{X}_{({\rm E})}^\nu
	= -u^\nu u_\alpha \, \mathcal{X}^{\mu\alpha}+u^\mu u_\alpha \, \mathcal{X}^{\nu\alpha}
	=
	\left( 
	\begin{array}{cccc}
	0    & \mathcal{X}_{01}  &  \mathcal{X}_{02} & \mathcal{X}_{03}\\
	-\mathcal{X}_{01} &  0   &  0 & 0 \\
	-\mathcal{X}_{02} & 0 &  0   & 0\\
	-\mathcal{X}_{03} & 0  & 0 & 0 
	\end{array} 
	\right)
,\\
\label{eq:Aperp}
	\Xperp^{\mu\nu}
	= \epsilon^{\alpha\beta\mu\nu} u_{\beta} \, \mathcal{X}_{({\rm M})\, \alpha}
	=\perp^{\mu\alpha}\perp^{\nu\beta}\mathcal{X}_{\alpha\beta}
	=\left( 
	\begin{array}{cccc}
	0    & 0  &  0 & 0\\
	0 &  0   &  \mathcal{X}_{12} & \mathcal{X}_{13} \\
	0& -\mathcal{X}_{12} &  0   & \mathcal{X}_{23} \\
	0& -\mathcal{X}_{13}  & -\mathcal{X}_{23} & 0 
	\end{array} 
	\right)
,
\end{gather}
where the matrix expressions are written in the comoving frame,
and the ``electric'' and ``magnetic'' four-vectors $\mathcal{X}^{\mu}_{({\rm E})}$ and $\mathcal{X}^{\mu}_{({\rm M})}$
are defined as [cf.\ Eqs.~\eqref{eq:Velectr}--\eqref{eq:Wmagn}]
\begin{gather}
\label{eq:Aelectr}
	\mathcal{X}^{\mu}_{({\rm E})} \equiv  
	u_\nu \mathcal{X}^{\mu\nu}
,\\
\label{eq:Amagn}
	\mathcal{X}^{\mu}_{({\rm M})} 
	\equiv
	\frac{1}{2} \, \epsilon^{\mu \nu \alpha \beta} \, u_{\nu} \, \mathcal{X}_{\alpha \beta}
.
\end{gather}
Note that the following relations are satisfied:
\begin{gather}
	\mathcal{X}^{\mu\nu} = \Xpar^{\mu\nu} + \Xperp^{\mu\nu}
,\\
	\perp_{\mu\nu} \Xpar^{\mu\nu} = 0
,\\
	u_\nu \Xperp^{\mu\nu} = 0
,
\end{gather}
and $\Xpar^{\mu\nu}$ and $\Xperp^{\mu\nu}$ can be expressed in terms of,
respectively,
``electric'' and ``magnetic'' four-vectors $\mathcal{X}^{\mu}_{({\rm E}i)}$ and
$\mathcal{X}^{\mu}_{({\rm M}i)}$
[see the first equalities in Eqs.\ (\ref{eq:Apar}) and (\ref{eq:Aperp})].
Similarly, the vortex contributions $\mathcal{T}^{\mu\nu}_{({\rm VE})}$ and $\mathcal{T}^{\mu\nu}_{({\rm VM})}$
to the energy-momentum tensor
can be presented as [see equations (88) and (89) in GD16]
\begin{eqnarray}
\label{eq:TVE}
	\mathcal{T}^{\mu\nu}_{({\rm VE})} &=&  
	\perp^{\mu\nu} \mathcal{W}_{({\rm E}i)}^{\alpha}\mathcal{V}_{({\rm E}i)\alpha }
	-\mathcal{W}_{({\rm E}i)}^{\mu} \mathcal{V}^{\nu}_{({\rm E}i)}
,\\
\label{eq:TVM}
	\mathcal{T}^{\mu\nu}_{({\rm VM})} &=& 
	\Wperp^{\mu\alpha}_{(i)}\Vperp^{\nu}_{(i)\,\alpha} + u^{\nu} \Wperp^{\mu\alpha}_{(i)} \mathcal{V}_{({\rm E}i)\alpha}
	+u^{\mu}\Wperp^{\nu\alpha}_{(i)}\mathcal{V}_{({\rm E}i)\alpha}
.
\end{eqnarray}

To sum up, the
dissipative equations governing dynamics of superfluid and superconducting mixture
consist of
the continuity equations \eqref{eq:continuity} [with $j_{(i)}^\mu$ given by Eq.~\eqref{eq:jmu}],
the energy-momentum conservation law \eqref{eq:dTmunu=0}
[with $T^{\mu\nu}$ given by Eqs.~\eqref{eq:Tmunu} and \eqref{eq:dT-EM-vortex}],
Maxwell equations \eqref{eq:maxwell-1} and \eqref{eq:maxwell-2},
and the superfluid equation [Eq.~\eqref{eq:Vmunu=0} or Eq.~\eqref{eq:sfl-eqn} below].
These equations are supplemented by the thermodynamic relations \eqref{eq:2ndlaw}, \eqref{eq:pres}, and \eqref{eq:dP},
as well as by the definition \eqref{eq:udtau=0} of the comoving frame. 

\section{Entropy generation equation}
\label{sec:entropy}

The equations of Sec.~\ref{sec:gen-eqns}
contain the entropy generation equation,
which is crucial for determining the general form of dissipative corrections
(see Sec.~\ref{sec:f-dj}).
One can derive this equation by considering the expression
$u_\nu \partial_{\mu} T^{\mu\nu} - u_\nu G^\nu$,
which vanishes in view of Eq.~\eqref{eq:dTmunu=0}.
Using Eqs.~\eqref{eq:continuity}, \eqref{eq:jmu},
\eqref{eq:uw=0},
\eqref{eq:Tmunu},
\eqref{eq:2ndlaw}, \eqref{eq:pres},
as well as the identities
$u_\nu \partial_\mu u^\nu = 0$, and $\partial_{\mu} g^{\mu\nu} = 0$,
we arrive at the following entropy generation equation
[cf.\ equation~(33) in Ref.~\cite{gusakov07}, equation~(58) in GD16, and equation~(25) in DGS20],
\begin{multline}
\label{eq:dSmu1}
\partial_{\mu} \left(S u^\mu \right)
= \frac{1}{T} u_\nu Y_{ik} w_{(k)\mu}
	\left[
		\widetilde{\mathcal{V}}^{\mu\nu}_{(i)}
		- \partial_\mu \left(\varkappa_i u_\nu \right)
		+ \partial_\nu \left(\varkappa_i u_\mu \right)
	\right]
+ \frac{\mu_i}{T} \partial_{\mu} \Delta j_{(i)}^\mu
- \frac{\mu_i}{T} \Delta\Gamma_i
\\- \frac{u^\mu}{T} \partial_{\mu} \varepsilon_{\rm add}
+ \frac{u_\nu}{T} \partial_{\mu} \left(
	\Delta T^{\mu\nu}_{({\rm EM+vortex})} + \Delta \tau^{\mu\nu}
	\right)
- \frac{Q}{T}
,
\end{multline}
where
\begin{gather}
\label{eq:tildeV}
	\widetilde{\mathcal{V}}^{\mu\nu}_{(i)}
	\equiv
	\mathcal{V}^{\mu\nu}_{(i)} - e_i F^{\mu\nu}
	=
	\partial^\mu\left[ w^\nu_{(i)}+\left(\mu_i + \varkappa_i\right) u^{\nu}\right]
		-\partial^\nu\left[ w^\mu_{(i)}+\left(\mu_i + \varkappa_i\right) u^{\mu}\right]
,
\end{gather}
and we
defined
$Q \equiv u_\nu G^\nu$.
Now, let us 
make use of 
Eqs.\ \eqref{eq:de-EM-vortex}
and \eqref{eq:dT-EM-vortex}
and substitute expressions
for
$d\varepsilon_{\rm add}$
and
$\Delta T^{\mu\nu}_{({\rm EM+vortex})}$.
Using equation (85) of GD16,
we present the term
$-u^{\mu} \partial_{\mu}\varepsilon_{\rm add}$ 
as
\begin{gather}
\label{eq:udeEM}
	-u^{\mu} \partial_{\mu}\varepsilon_{\rm add} =
	u^{\nu} F_{\mu\nu} \, \partial_{\alpha}\left( 
	\frac{1}{4\pi} G^{\mu\alpha}\right)
	+ u^{\nu} \mathcal{V}_{(i)\mu\nu} \, \, \partial_{\alpha} \mathcal{W}^{\mu\alpha}_{(i)}
	-  u_\nu \partial_\mu \Delta T^{\mu\nu}_{({\rm EM+vortex})}
.
\end{gather}
Then, employing Maxwell equations \eqref{eq:maxwell-2}
together with the relation $u_\mu \partial_\nu w_{(i)}^\mu = - w_{(i)}^\mu \partial_\nu u_\mu$ [which follows from Eq.~\eqref{eq:uw=0}]
and substituting 
Eqs.\ \eqref{eq:Jfree}, \eqref{eq:dT-EM-vortex}, \eqref{eq:tildeV}, and \eqref{eq:udeEM}
into Eq.~\eqref{eq:dSmu1}, we obtain%
\begin{gather}
\label{eq:dSmu-EM2}
	\partial_{\mu} S^\mu
	=
	\frac{n_i}{T} 
		u^{\nu} \, \mathcal{V}_{(i)\mu\nu}
		W_{(i)}^\mu
	- \Delta j_{(i)}^\mu 
		d_{(i)\mu}
	- \varkappa_i \nablaperp_\mu \left( \frac{Y_{ik} w^{\mu}_{(k)}}{T} \right)
	- \Delta \tau^{\mu\nu} \partial_{\mu} \left(\frac{u_\nu}{T}\right) 
	- \frac{\mu_i}{T} \Delta\Gamma_i
	- \frac{Q}{T}
,
\end{gather}
where we introduced the entropy four-current
\begin{gather}
\label{eq:Smu}
	S^\mu = S u^\mu
		- \frac{\mu_i}{T} \Delta j_{(i)}^\mu
		- \frac{\varkappa_i}{T}  Y_{ik} w^{\mu}_{(k)}
		- \frac{u_\nu}{T} \Delta \tau^{\mu\nu}
,
\end{gather}
the four-vector $W^{\mu}_{(i)}$,
\begin{gather}
\label{eq:Wmu}
	W^{\mu}_{(i)}
	\equiv \frac{1}{n_i}\left[
		Y_{ik} w^\mu_{(k)} + \perp^{\mu\nu} \partial^{\alpha} 
	\mathcal{W}_{\nu \alpha (i)} 
	\right]
,
\end{gather}
the displacement vector (see DGS20)
\begin{gather}
\label{eq:dmu}
	d_{(i)\mu} \equiv
	\nablaperp_{\mu} \left(\frac{\mu_i}{T}\right) 
					- \frac{e_i E_\mu}{T}
,
\end{gather}
and the orthogonal part of the four-gradient
\begin{gather}
\label{eq:nabla}
	\nablaperp_{\mu} \equiv \perp_{\mu\nu} \partial^\nu	
.
\end{gather}
Note that $d_{(i)\mu}$ and $W^{\mu}_{(i)}$ can be defined up to an arbitrary term proportional to $u^\mu$,
which does not affect the entropy generation equation \eqref{eq:dSmu-EM2}
due to the condition \eqref{eq:udj=0} and antisymmetry of $\mathcal{V}^{\mu\nu}_{(i)}$, respectively. 
For further convenience, we define these vectors in a way that ensures that
they are both orthogonal to $u^\mu$.\footnote{\label{footnote:Wmu}
GD16
uses a slightly different definition for $W^{\mu}_{(i)}$:
$W^{\mu}_{(i)}
	\equiv (1/n_i)\left[
		Y_{ik} w^\mu_{(k)} + \partial_{\alpha} 
	\mathcal{W}^{\mu \alpha}_{(i)} 
	\right]$.
If one prefers to use that definition, then one should replace $W^{\mu}_{(i)}$ with $\perp^{\mu\nu} W_{\nu(i)}$
[which is equivalent to Eq.~\eqref{eq:Wmu} due to the condition \eqref{eq:uw=0}]
everywhere in
the paper.
}

If $u^\mu$ is
specified by the condition \eqref{eq:udtau=0},
Eqs.~\eqref{eq:dSmu-EM2} and \eqref{eq:Smu} reduce to\footnote{
As in DGS20, we make use of the condition \eqref{eq:udtau=0} and replace
$\Delta \tau^{\mu\nu} \partial_{\mu} \left({u_\nu}/{T}\right)$
with $\Delta \tau^{\mu\nu} ({\nablaperp_{\mu} u_\nu})/{T}$
in the right-hand side of Eq.~\eqref{eq:dSmu-EM-LL}.
}
\begin{gather}
\label{eq:dSmu-EM-LL}
	\partial_{\mu} S^\mu
	= 	
		\frac{\mu_i n_i^2}{T}
	    	f_{(i)\mu} W^{\mu}_{(i)}
		- \Delta j_{(i)}^\mu d_{(i)\mu}
    	- \varkappa_i \nablaperp_\mu \left( \frac{Y_{ik} w^{\mu}_{(k)}}{T} \right)
		- \Delta \tau^{\mu\nu} \frac{\nablaperp_{\mu} u_\nu}{T} 
		- \frac{\mu_i}{T} \Delta\Gamma_i
		- \frac{Q}{T}
,\\
\label{eq:Smu-LL}
	S^\mu = S u^\mu
		- \frac{\mu_i}{T} \Delta j_{(i)}^\mu
		- \frac{\varkappa_i}{T}  Y_{ik} w^{\mu}_{(k)}
.
\end{gather}
Here we
introduced
the four-vector $f_{(i)}^\mu$ as
\begin{gather}
\label{eq:sfl-eqn}
	u_{\nu} \mathcal{V}_{(i)}^{\mu\nu}
	= \mu_i n_i f_{(i)}^\mu
,
\end{gather}
where no summation over repeated index $i$ is assumed.
Note that $f_{(i)}^\mu$ is orthogonal to $u^\mu$, since the vorticity tensor $\mathcal{V}_{(i)}^{\mu\nu}$ is antisymmetric,
\begin{gather}
\label{eq:uf=0}
	u_\mu f_{(i)}^\mu = 0
.
\end{gather}
Eq.~\eqref{eq:sfl-eqn} can be regarded as a superfluid equation \cite{gusakov16,gd16},
which replaces the potentiality condition $\mathcal{V}_{(i)}^{\mu\nu} = 0$ of a vortex-free system.

The right-hand side of Eq.~\eqref{eq:dSmu-EM-LL}
describes entropy generation and must be non-negative (except for the arbitrary last term)
for all possible fluid configurations.
It includes
vortex-mediated mutual friction between normal and superfluid components (first term) \cite{hs18},
diffusion (second term),
viscosity (third and fourth terms),
chemical reactions (such as Urca-processes; fifth term)
and radiation (sixth term).

    Note, in passing, that different formulations of the first-order hydrodynamics 
    (i.e., different forms
    of dissipative corrections)
    are possible
    even if $u^\mu$ is specified unambiguously \cite{kovtun19}.
    This is due to the fact that various derivatives
    entering the dissipative corrections
    are not all independent, but can be expressed (up to higher-order terms) through one another 
    using
    the 
    zero-order (nondissipative) hydrodynamic equations.
    For example, one can relate
    $u^\nu \partial_\nu u^\mu$
    to $\nablaperp^\mu P$
    via the momentum conservation law
    $\nablaperp_\nu T^{\mu\nu} = 0$.
    We follow here the approach of Ref.\ \cite{ll87}, so that in our formulation the right-hand side of Eq.\ \eqref{eq:dSmu-EM-LL} (and, consequently, the dissipative corrections) in the comoving frame
    contain
    only spatial derivatives and 
    do not
    contain the terms like 
    $u^\nu \partial_\nu u^\mu$ or $u^\nu \partial_\nu T$.

\section{Diffusive currents and mutual friction forces}
\label{sec:f-dj}

The entropy generation equation \eqref{eq:dSmu-EM-LL} allows one to find
the general form of the unknown dissipative corrections,%
\footnote{Note that
	some of	these corrections
	may, in fact,
	contain nondissipative terms,
	but, for brevity, we
	call them `dissipative'.
}
namely $f_{(i)}^\mu$, $\Delta j_{(i)}^\mu$, $\varkappa_i$, $\Delta \tau^{\mu\nu}$, and $\Delta \Gamma_i$
(here and below we ignore the last term, $-Q/T$, which can
be arbitrary).
Following Landau \& Lifshitz \cite{ll87} and DGS20,
we express the dissipative corrections
as linear combinations of thermodynamic forces
$W^{\mu}_{(i)}$,
$d_{(i)\mu}$,
$\nablaperp_\mu \left( {Y_{ik} w^{\mu}_{(k)}} / {T} \right)$,
$\nablaperp_{\mu} u_\nu$,
and $\mu_i$,%
\footnote{
Actually,
$\mu_i$ should enter these expressions
	only
	in particular combinations that represent chemical potential imbalances for a given reaction
	(e.g., $\mu_n - \mu_p - \mu_e$ for the direct or modified Urca processes \cite{ykgh01});
	see DGS20 for more details.
}
and require
that the right-hand side of Eq.~\eqref{eq:dSmu-EM-LL}
would be a positively defined quadratic form,
so that 
the entropy
would not decrease for all possible fluid configurations.
The coefficients
arising in these linear combinations
can be scalars, vectors, or tensors, that can only depend on the system properties in the absence of dissipation;
they are collectively called {\it transport coefficients}.
We require, in addition,
that these coefficients
must satisfy the Onsager
relations.

In the completely isotropic (in the comoving frame) matter, 
the transport coefficients
depend only on the equilibrium scalar thermodynamic quantities,
as well as on $u^\mu$ and $g^{\mu\nu}$ (or $\perp^{\mu\nu} \equiv g^{\mu\nu}+u^{\mu} u^{\nu}$).
In the presence of preferred directions (e.g., vortex lines or magnetic field),
the coefficients, generally, depend also on the corresponding vectors
and the angles between them.
These vectors include
superfluid vectors $w_{(i)}^\mu$,
electromagnetic vectors $E^\mu$, $D^\mu$, $B^\mu$, and $H^\mu$,
and vortex-related vectors
$\mathcal{V}_{({\rm E}i)}^{\mu}$, $\mathcal{V}_{({\rm M}i)}^{\mu}$,
$\mathcal{W}_{({\rm E}i)}^{\mu}$, and $\mathcal{W}_{({\rm M}i)}^{\mu}$.
However, the situation is considerably simplified
in the MHD approximation described in Sec.~\ref{sec:mhd-limit} (see also GD16).
This approximation is mainly based on the fact that 
the magnetic induction ${\pmb B}$ is much larger than the fields 
${\pmb E}$, ${\pmb D}$, and ${\pmb H}$ in the comoving frame,
and is locked to superconducting proton
flux tubes.
In this limit
the only preferred directions%
\footnote{
That these preferred directions are the only ones that should be taken
into account 
in the MHD approximation 
is independently justified by the results of Appendix B,
where it is shown that 
the more microscopic approach leads
to exactly the same 
dissipative corrections
as those
obtained in this section.
Generally, any additional preferred direction can be ignored
as long as one can neglect the corresponding force
in the force balance equations for particles
or vortices.
For example, in the nonsuperfluid MHD
in the limit ${\pmb B} \to 0$
an anisotropic correction to the diffusion coefficients $\mathcal{D}_{e\mu}^{\mu\nu}$
is of the order 
$\sim (e_p n_i B)/(c J_{e\mu}) 
\sim \text{(Lorentz force)} / (e\mu \text{ friction force})$,
see DGS20.
Correspondingly, the magnetic field does not provide
a preferred direction in this limit.
}
are defined
by the neutron vortex lines $\mathcal{V}_{({\rm M}n)}^\mu$,
proton vortex lines $\mathcal{V}_{({\rm M}p)}^\mu$
[or, equivalently, the magnetic induction $B^\mu$, see Eq.~\eqref{eq:mhd-limit:VMp}],
and the 
superfluid neutron current,
$Y_{nk} w_{(k)}^\mu$.%
\footnote{
In the thermodynamic equilibrium,
  the superconducting proton current $Y_{pk} w_{(k)}^\mu$ vanishes 
  in the MHD approximation
  due to the screening condition [see Eq.\ \eqref{eq:full-mhd:screening} with $\Delta j^\mu_{(i)}=0$].
}
Below,
following Refs.\ \cite{khalatnikov00, putterman74, ll87},
we neglect small terms  that explicitly depend on $w_{(k)}^\mu$
(or, equivalently, on $Y_{nk} w_{(k)}^\mu$)
{\it in the expressions for the transport coefficients}.
These terms are usually ignored in the literature
\cite{khalatnikov00, putterman74, ll87}
when deriving the dissipative hydrodynamic equations for superfluid helium-4.
In the context of neutron stars, the same approximation has been adopted and discussed in Ref.\ \cite{gusakov07}.
As a result, we are left with only two preferred directions, specified by
the neutron vortices $\mathcal{V}_{({\rm M}n)}^\mu$ 
and magnetic field/proton flux tubes $B^\mu$ [or $\mathcal{V}_{({\rm M}p)}^\mu$],
which determine anisotropy of transport coefficients.

Under the above assumptions,
the vectors $\Delta j_{(i)}^\mu$ and $f_{(i)}^\mu$
can only depend on the thermodynamic forces
$W^{\mu}_{(i)}$ and $d_{(i)}^{\mu}$ 
(and are independent of the forces $\nablaperp_\mu \left( {Y_{ik} w^{\mu}_{(k)}} / {T} \right)$,
$\nablaperp_{\mu} u_\nu$,
and $\mu_i$):%
\footnote{
	See
	Appendix B of DGS20,
	where it is demonstrated,
	for a similar problem,
	that $\Delta j_{(i)}^\mu$ cannot depend on the tensor $\nablaperp_{\mu} u_\nu$.
	The same consideration also applies 
	to $f_{(i)}^\mu$ and can be readily generalized
	to an arbitrary number of preferred axial vectors
	(such as $\mathcal{V}_{({\rm M}n)}^\mu$ and/or $\mathcal{V}_{({\rm M}p)}^\mu$)
	in the system.
	In turn, it
	is also easy to verify that $\Delta j_{(i)}^\mu$ and $f_{(i)}^\mu$
	cannot depend on the scalar thermodynamic forces,
	such as $\nablaperp_\nu \left( {Y_{ik} w^{\nu}_{(k)}} / {T} \right)$.
	This dependence may only lead to additional terms
	$\propto u^\mu \nablaperp_\nu \left( {Y_{ik} w^{\nu}_{(k)}} / {T} \right)$
	in Eqs.~\eqref{eq:general:f} and \eqref{eq:general:dj},
	but these terms must vanish 
	to satisfy the
	conditions
	\eqref{eq:udj=0}
	and
	\eqref{eq:uf=0}.
}
\begin{gather}
\label{eq:general:f}
	- \frac{\mu_i n_i^2}{T} f_{(i)}^\mu	= - \mathcal{A}_{ik}^{\mu\nu} W_{(k)\nu} -  \mathcal{B}_{ik}^{\mu\nu} d_{(k)\nu}
,\\
\label{eq:general:dj}
	\Delta j_{(i)}^\mu	= - \mathcal{C}_{ik}^{\mu\nu} W_{(k)\nu} - \mathcal{D}_{ik}^{\mu\nu} d_{(k)\nu}
,
\end{gather}
where {\it no} summation over $i$ in the left-hand side of Eq.~\eqref{eq:general:f} is implied.
The transport coefficient $\mathcal{A}_{ik}^{\mu\nu}$
describes the mutual friction effects \cite{hs18},
as well as (possible)
interaction between neutron vortices and proton flux tubes.%
\footnote{
	Note that the
	vortex-flux tube
	interaction should be accounted for
	in the expressions for $\mathcal{W}_{({\rm M}i)}^\mu$
	[which enter
	the definition \eqref{eq:Wmu} for $W_{(i)}^{\mu}$].
	In Sec.~\ref{sec:mhd-limit} we employ a simple model
	which ignores this effect [see Eq.~\eqref{eq:mhd-limit:WMi}];
	however, 
	such a simplification does not affect the general expression \eqref{eq:oB:Aik-2} for the coefficient $\mathcal{A}_{ik}^{\mu\nu}$.
}
The coefficient $\mathcal{D}_{ik}^{\mu\nu}$
is responsible for
the diffusion, thermodiffusion and thermal conductivity effects (see DGS20).
Finally, the cross-coefficients $\mathcal{B}_{ik}^{\mu\nu}$ and $\mathcal{C}_{ik}^{\mu\nu}$
describe the impact of diffusive currents on the mutual friction forces on vortices, and vice versa.

\newpage
In the present work, we are mainly interested in studying
the joint effects of diffusion and vortices
(represented by the vectors $\Delta j_{(i)}^\mu$ and $f_{(i)}^\mu$)
on the structure of superfluid MHD.
To study these effects, it is sufficient to consider only the first two terms in Eq.~\eqref{eq:dSmu-EM-LL},
since they do not interfere with the other terms in this equation
and constitute a positively defined quadratic form themselves
[see Eqs.\ (\ref{eq:general:f}) and (\ref{eq:general:dj})].
Thus, in what follows, we shall ignore viscosity ($\varkappa_i = \Delta \tau^{\mu\nu} = 0$) and chemical reactions ($\Delta \Gamma_i = 0$): the related dissipative corrections can be studied separately and, in fact,
have already been analyzed in the past (see, e.g., Refs.\ \cite{gusakov07,adhc17,rw20} and DGS20).
With this simplification, the entropy generation equation \eqref{eq:dSmu-EM-LL}
becomes
\begin{gather}
\label{eq:dSmu-diff-mf}
\partial_{\mu} 
	\left(
		S u^\mu	
		- \frac{\mu_i}{T} \Delta j_{(i)}^\mu
	\right)
= 	\frac{\mu_i n_i^2}{T}
f_{(i)\mu} W^{\mu}_{(i)}
- \Delta j_{(i)}^\mu d_{(i)\mu}
.
\end{gather}

The coefficients
$\mathcal{A}_{ik}^{\mu\nu}$, $\mathcal{B}_{ik}^{\mu\nu}$, $\mathcal{C}_{ik}^{\mu\nu}$, and $\mathcal{D}_{ik}^{\mu\nu}$ 
in Eqs.~\eqref{eq:general:f} and \eqref{eq:general:dj}
depend on the vectors $\mathcal{V}_{({\rm M}n)}^\mu$
and $B^\mu$, as well as on the scalar thermodynamic quantities
and on $u^{\mu}$ and $\perp^{\mu\nu}$.
Below we
provide expressions for
these coefficients
for the system with two preferred directions
and demonstrate how these expressions can be simplified
in the case of
only one preferred direction.

\subsection{General case: two preferred directions}
\label{sec:f-dj:oB}

Let us introduce the following quantities:
\begin{gather}
b^\mu \equiv \frac{B^\mu}{\sqrt{B_\alpha B^\alpha}}
,\\
b^{\mu\nu} \equiv \frac{\Fperp^{\mu\nu}}{\sqrt{B_\alpha B^\alpha}}
,\\
\omega^\mu \equiv \frac{\mathcal{V}_{({\rm M}n)}^\mu}{\sqrt{\mathcal{V}_{({\rm M}n)\alpha} \mathcal{V}_{({\rm M}n)}^\alpha}}
,\\
\omega^{\mu\nu} \equiv \frac{\Vperp_{({\rm M}n)}^{\mu\nu}}{\sqrt{\mathcal{V}_{({\rm M}n)\alpha} \mathcal{V}_{({\rm M}n)}^\alpha}}
.
\end{gather}
In the comoving frame $b^\mu = (0, {\pmb b})$, $\omega^\mu = (0, {\pmb \omega})$,
where ${\pmb b}$ and ${\pmb \omega}$ are the unit vectors in the direction of the magnetic field
and neutron vortices, respectively.

Onsager principle 
leads to conditions%
\footnote{%
The minus sign in Eq.~\eqref{eq:oB:Cik=Bki} 
appears because $d_{(k)\nu}$ and $W_{(k)\nu}$ have different parity under time reversal $t \to -t$ \cite{ll5}.
}
\begin{gather}
\label{eq:oB:Aik=Aki}
	\mathcal{A}_{ik}^{\mu\nu}({\pmb b}, {\pmb \omega}) = \mathcal{A}_{ki}^{\nu\mu} (-{\pmb b}, -{\pmb \omega})
,\\
\label{eq:oB:Dik=Dki}
	\mathcal{D}_{ik}^{\mu\nu}({\pmb b}, {\pmb \omega}) = \mathcal{D}_{ki}^{\nu\mu} (-{\pmb b}, -{\pmb \omega})
,\\
\label{eq:oB:Cik=Bki}
	\mathcal{C}_{ik}^{\mu\nu}({\pmb b}, {\pmb \omega}) = - \mathcal{B}_{ki}^{\nu\mu}(-{\pmb b}, -{\pmb \omega})
.
\end{gather}
From the constraints $u_\mu f_{(i)}^\mu = 0$ [Eq.\ \eqref{eq:uf=0}] and $u_\mu \Delta j_{(i)}^\mu = 0$ [Eq.\ \eqref{eq:udj=0}]
it also follows that
\begin{gather}
\label{eq:uA=0}
	u_\mu \mathcal{A}_{ik}^{\mu\nu} = u_\mu \mathcal{B}_{ik}^{\mu\nu} = u_\mu \mathcal{C}_{ik}^{\mu\nu} = u_\mu \mathcal{D}_{ik}^{\mu\nu} = 0
.
\end{gather}
Relations \eqref{eq:oB:Aik=Aki}--\eqref{eq:uA=0} imply that
all transport coefficients 
are purely spatial in the comoving frame
and may depend on $u^\mu$ only through the tensor $\perp^{\mu\nu}$.

Let us start with the
transport
coefficient $\mathcal{A}_{ik}^{\mu\nu}$.
Generally, it can be presented as a sum of nine linearly independent tensors\footnote{
To make this point clearer,
let us work in the comoving frame, choosing $x$-axis along the direction ${\pmb \omega}$
and $z$ axis along $\left[ {\pmb \omega} \times {\pmb b} \right]$.
Then, introducing unit vectors
$y^\mu \equiv \frac{b^\mu - b^\alpha \omega_\alpha \omega^\mu}{|| b^\nu - b^\alpha \omega_\alpha \omega^\nu ||} = (0,0,1,0)$
and 
$z^\mu \equiv - y_\alpha \omega^{\mu\alpha} = (0,0,0,1)$,
one can generally decompose
$\mathcal{A}_{ik}^{\mu\nu}$ 
into the
sum of nine linearly independent tensors,
\begin{gather*}
\begin{split}
\mathcal{A}_{ik}^{\mu\nu}
=&~\mathcal{A}_{ik}^{11} \omega^\mu \omega^\nu
+ \mathcal{A}_{ik}^{12} \omega^\mu y^\nu
+ \mathcal{A}_{ik}^{13} \omega^\mu z^\nu
+ \mathcal{A}_{ik}^{21} y^\mu \omega^\nu
+ \mathcal{A}_{ik}^{22} y^\mu y^\nu
+ \mathcal{A}_{ik}^{23} y^\mu z^\nu
+ \mathcal{A}_{ik}^{31} z^\mu \omega^\nu
+ \mathcal{A}_{ik}^{32} z^\mu y^\nu
+ \mathcal{A}_{ik}^{33} z^\mu z^\nu
,
\end{split}
\end{gather*}
where the scalar coefficients
$\mathcal{A}_{ik}^{11}$, $\mathcal{A}_{ik}^{12}$ \ldots may depend on the angle between ${\pmb \omega}$ and ${\pmb b}$.
One can directly check that the nine tensors entering Eq.~\eqref{eq:oB:Aik-2}
are indeed linearly independent
and they can be expressed as linear combinations of $\omega^\mu \omega^\nu$, $\omega^\mu y^\nu$, $\omega^\mu z^\nu$, etc.
},
which we choose in the following form that allows us
to separate symmetric (the first six terms)
and antisymmetric (the last three terms) parts of the tensor:
\begin{gather}
\label{eq:oB:Aik-2}
\begin{split}
	\mathcal{A}_{ik}^{\mu\nu}
	=&~\mathcal{A}_{ik}^{\perp} \perp^{\mu\nu}
	+ \mathcal{A}_{ik}^{\omega\omega} \omega^\mu \omega^\nu
	+ \mathcal{A}_{ik}^{bb} b^\mu b^\nu
	\\&
	+ \mathcal{A}_{ik}^{\omega b}  (\omega^\mu b^\nu + \omega^\nu b^\mu)
	+ \mathcal{A}_{ik}^{\omega \omega b}  (\omega^\mu \omega_\alpha b^{\nu\alpha} + \omega^\nu \omega_\alpha b^{\mu\alpha})
	+ \mathcal{A}_{ik}^{b \omega b}  (b^\mu \omega_\alpha b^{\nu\alpha} + b^\nu \omega_\alpha b^{\mu\alpha})
	\\&
	+ \mathcal{A}_{ik}^{\omega-b}(\omega^\mu b^\nu - \omega^\nu b^\mu)
	+ \mathcal{A}_{ik}^{\omega} \omega^{\mu\nu}
	+ \mathcal{A}_{ik}^{b} b^{\mu\nu}
,
\end{split}
\end{gather}
where the scalar coefficients $\mathcal{A}_{ik}^{\perp}$, $\mathcal{A}_{ik}^{\omega\omega}$ etc. 
may depend on the equilibrium quantities and the angle between ${\pmb b}$ and ${\pmb \omega}$.
To clarify the meaning of different terms in Eq.~\eqref{eq:oB:Aik-2}, it is instructive
to write out the expression for the vector $\mathcal{A}_{ik}^{\mu\nu} W_{(k)\nu}$ in the comoving frame.
The zeroth component of this four-vector vanishes, 
while its spatial part reads
\begin{gather}
\label{eq:oB:Aik-Wk}
\begin{split}
	&\mathcal{A}_{ik}^{\perp} {\pmb W}_k
		+ \mathcal{A}_{ik}^{\omega\omega} {\pmb \omega} \left({\pmb \omega} {\pmb W}_k\right)
		+ \mathcal{A}_{ik}^{bb} {\pmb b} \left({\pmb b} {\pmb W}_k\right)
		\\&
		+ \mathcal{A}_{ik}^{\omega b} 
			\left[
				{\pmb \omega} \left({\pmb b} {\pmb W}_k\right)
				+ {\pmb b} \left({\pmb \omega} {\pmb W}_k\right)
			\right]  
		+ \mathcal{A}_{ik}^{\omega \omega b} 
			\left\{
				{\pmb \omega} \left( \left[{\pmb \omega} \times {\pmb b}\right] {\pmb W}_k\right)
				+ \left[{\pmb \omega} \times {\pmb b}\right] \left( {\pmb \omega} {\pmb W}_k\right)
			\right\}
		+ \mathcal{A}_{ik}^{b \omega b} 
			\left\{
				{\pmb b} \left( \left[{\pmb \omega} \times {\pmb b}\right] {\pmb W}_k\right)
				+ \left[{\pmb \omega} \times {\pmb b}\right] \left( {\pmb b} {\pmb W}_k\right)
			\right\}
		\\&
		+ \mathcal{A}_{ik}^{\omega-b}
			\left[
				{\pmb \omega} \left({\pmb b} {\pmb W}_k\right)
				- {\pmb b} \left({\pmb \omega} {\pmb W}_k\right)
			\right]
		+ \mathcal{A}_{ik}^{\omega} \left[ {\pmb W}_k \times {\pmb \omega} \right] 
		+ \mathcal{A}_{ik}^{b} \left[ {\pmb W}_k \times {\pmb b} \right] 
,
\end{split}
\end{gather}
where ${\pmb W}_k$ is the spatial part of the four-vector $W_{(k)}^\mu$: $W_{(k)}^\mu = (0, {\pmb W}_k)$.

Plugging Eq.~\eqref{eq:oB:Aik-2} into the Onsager relation \eqref{eq:oB:Aik=Aki}, we get
\begin{gather}
\label{eq:oB:Aik=Aki-2}
\begin{split}
	&\mathcal{A}_{ik}^{\perp} = \mathcal{A}_{ki}^{\perp}
,\quad
	\mathcal{A}_{ik}^{\omega\omega} = \mathcal{A}_{ki}^{\omega\omega}
,\quad
	\mathcal{A}_{ik}^{bb} = \mathcal{A}_{ki}^{bb}
,\\
	&\mathcal{A}_{ik}^{\omega b} = \mathcal{A}_{ki}^{\omega b}
,\quad
	\mathcal{A}_{ik}^{\omega \omega b} = - \mathcal{A}_{ki}^{\omega \omega b}  
,\quad
	\mathcal{A}_{ik}^{b \omega b} = - \mathcal{A}_{ki}^{b \omega b}  
\\
	&\mathcal{A}_{ik}^{\omega-b} = - \mathcal{A}_{ki}^{\omega-b}
,\quad
	\mathcal{A}_{ik}^{\omega} = \mathcal{A}_{ki}^{\omega} 
,\quad
	\mathcal{A}_{ik}^{b} = \mathcal{A}_{ki}^{b} 
.
\end{split}
\end{gather}
As one can check by substituting Eqs.\ \eqref{eq:general:f}, \eqref{eq:oB:Aik-2}, and \eqref{eq:oB:Aik=Aki-2} into the entropy generation equation \eqref{eq:dSmu-diff-mf},
the coefficients
$\mathcal{A}_{ik}^{\omega \omega b}$,
$\mathcal{A}_{ik}^{b \omega b} $,
$\mathcal{A}_{ik}^{\omega}$,
and $\mathcal{A}_{ik}^{b}$
are nondissipative and do not contribute to the entropy generation.

The same consideration also applies to the transport coefficients
$\mathcal{B}_{ik}^{\mu\nu}$, $\mathcal{C}_{ik}^{\mu\nu}$, and $\mathcal{D}_{ik}^{\mu\nu}$.
The result is
\begin{gather}
\label{eq:oB:Bik}
\begin{split}
\mathcal{B}_{ik}^{\mu\nu}
=&~\mathcal{B}_{ik}^{\perp} \perp^{\mu\nu}
+ \mathcal{B}_{ik}^{\omega\omega} \omega^\mu \omega^\nu
+ \mathcal{B}_{ik}^{bb} b^\mu b^\nu
\\&
+ \mathcal{B}_{ik}^{\omega b}  (\omega^\mu b^\nu + \omega^\nu b^\mu)
+ \mathcal{B}_{ik}^{\omega \omega b}  (\omega^\mu \omega_\alpha b^{\nu\alpha} + \omega^\nu \omega_\alpha b^{\mu\alpha})
+ \mathcal{B}_{ik}^{b \omega b}  (b^\mu \omega_\alpha b^{\nu\alpha} + b^\nu \omega_\alpha b^{\mu\alpha})
\\&
+ \mathcal{B}_{ik}^{\omega-b}(\omega^\mu b^\nu - \omega^\nu b^\mu)
+ \mathcal{B}_{ik}^{\omega} \omega^{\mu\nu}
+ \mathcal{B}_{ik}^{b} b^{\mu\nu}
,
\end{split}
\\
\label{eq:oB:Cik}
\begin{split}
\mathcal{C}_{ik}^{\mu\nu}
=&~\mathcal{C}_{ik}^{\perp} \perp^{\mu\nu}
+ \mathcal{C}_{ik}^{\omega\omega} \omega^\mu \omega^\nu
+ \mathcal{C}_{ik}^{bb} b^\mu b^\nu
\\&
+ \mathcal{C}_{ik}^{\omega b}  (\omega^\mu b^\nu + \omega^\nu b^\mu)
+ \mathcal{C}_{ik}^{\omega \omega b}  (\omega^\mu \omega_\alpha b^{\nu\alpha} + \omega^\nu \omega_\alpha b^{\mu\alpha})
+ \mathcal{C}_{ik}^{b \omega b}  (b^\mu \omega_\alpha b^{\nu\alpha} + b^\nu \omega_\alpha b^{\mu\alpha})
\\&
+ \mathcal{C}_{ik}^{\omega-b}(\omega^\mu b^\nu - \omega^\nu b^\mu)
+ \mathcal{C}_{ik}^{\omega} \omega^{\mu\nu}
+ \mathcal{C}_{ik}^{b} b^{\mu\nu}
,
\end{split}
\\
\label{eq:oB:Dik}
\begin{split}
\mathcal{D}_{ik}^{\mu\nu}
=&~\mathcal{D}_{ik}^{\perp} \perp^{\mu\nu}
+ \mathcal{D}_{ik}^{\omega\omega} \omega^\mu \omega^\nu
+ \mathcal{D}_{ik}^{bb} b^\mu b^\nu
\\&
+ \mathcal{D}_{ik}^{\omega b}  (\omega^\mu b^\nu + \omega^\nu b^\mu)
+ \mathcal{D}_{ik}^{\omega \omega b}  (\omega^\mu \omega_\alpha b^{\nu\alpha} + \omega^\nu \omega_\alpha b^{\mu\alpha})
+ \mathcal{D}_{ik}^{b \omega b}  (b^\mu \omega_\alpha b^{\nu\alpha} + b^\nu \omega_\alpha b^{\mu\alpha})
\\&
+ \mathcal{D}_{ik}^{\omega-b}(\omega^\mu b^\nu - \omega^\nu b^\mu)
+ \mathcal{D}_{ik}^{\omega} \omega^{\mu\nu}
+ \mathcal{D}_{ik}^{b} b^{\mu\nu}
.
\end{split}
\end{gather}
The Onsager principle for $\mathcal{B}_{ik}^{\mu\nu}$ and $\mathcal{C}_{ik}^{\mu\nu}$ \eqref{eq:oB:Cik=Bki} implies
\begin{gather}
\label{eq:oB:Cik=Bki-2}
\begin{split}
&\mathcal{C}_{ik}^{\perp} = - \mathcal{B}_{ki}^{\perp}
,\quad
\mathcal{C}_{ik}^{\omega\omega} = - \mathcal{B}_{ki}^{\omega\omega}
,\quad
\mathcal{C}_{ik}^{bb} = - \mathcal{B}_{ki}^{bb}
,\\
&\mathcal{C}_{ik}^{\omega b} = - \mathcal{B}_{ki}^{\omega b}
,\quad
\mathcal{C}_{ik}^{\omega \omega b} = \mathcal{B}_{ki}^{\omega \omega b}  
,\quad
\mathcal{C}_{ik}^{b \omega b} = \mathcal{B}_{ki}^{b \omega b}  
\\
&\mathcal{C}_{ik}^{\omega-b} = \mathcal{B}_{ki}^{\omega-b}
,\quad
\mathcal{C}_{ik}^{\omega} = - \mathcal{B}_{ki}^{\omega} 
,\quad
\mathcal{C}_{ik}^{b} = - \mathcal{B}_{ki}^{b} 
.
\end{split}
\end{gather}
Note that the coefficients
$\mathcal{B}_{ik}^{\perp}$,
$\mathcal{B}_{ik}^{\omega\omega}$,
$\mathcal{B}_{ik}^{bb}$,
$\mathcal{B}_{ik}^{\omega b}$,
and $\mathcal{B}_{ik}^{\omega - b}$
are nondissipative,
in contrast to
to the analogous coefficients
$\mathcal{A}_{ik}^{\perp}$,
$\mathcal{A}_{ik}^{\omega\omega}$,
$\mathcal{A}_{ik}^{bb}$,
$\mathcal{A}_{ik}^{\omega b}$,
and $\mathcal{A}_{ik}^{\omega - b}$.

The Onsager principle for $\mathcal{D}_{ik}^{\mu\nu}$ \eqref{eq:oB:Dik=Dki} leads to
\begin{gather}
\label{eq:oB:Dik=Dki-2}
\begin{split}
&\mathcal{D}_{ik}^{\perp} = \mathcal{D}_{ki}^{\perp}
,\quad
\mathcal{D}_{ik}^{\omega\omega} = \mathcal{D}_{ki}^{\omega\omega}
,\quad
\mathcal{D}_{ik}^{bb} = \mathcal{D}_{ki}^{bb}
,\\
&\mathcal{D}_{ik}^{\omega b} = \mathcal{D}_{ki}^{\omega b}
,\quad
\mathcal{D}_{ik}^{\omega \omega b} = - \mathcal{D}_{ki}^{\omega \omega b}  
,\quad
\mathcal{D}_{ik}^{b \omega b} = - \mathcal{D}_{ki}^{b \omega b}  
\\
&\mathcal{D}_{ik}^{\omega-b} = - \mathcal{D}_{ki}^{\omega-b}
,\quad
\mathcal{D}_{ik}^{\omega} = \mathcal{D}_{ki}^{\omega} 
,\quad
\mathcal{D}_{ik}^{b} = \mathcal{D}_{ki}^{b} 
.
\end{split}
\end{gather}
The coefficients
$\mathcal{D}_{ik}^{\omega \omega b}$,
$\mathcal{D}_{ik}^{b \omega b} $,
$\mathcal{D}_{ik}^{\omega}$,
and $\mathcal{D}_{ik}^{b}$ are nondissipative,
similarly to
$\mathcal{A}_{ik}^{\omega \omega b}$,
$\mathcal{A}_{ik}^{b \omega b} $,
$\mathcal{A}_{ik}^{\omega}$,
and $\mathcal{A}_{ik}^{b}$.

In this section we have derived the general
expressions for
the transport coefficients
$\mathcal{A}_{ik}^{\mu\nu}$ \eqref{eq:oB:Aik-2},
$\mathcal{B}_{ik}^{\mu\nu}$ \eqref{eq:oB:Bik},
$\mathcal{C}_{ik}^{\mu\nu}$ \eqref{eq:oB:Cik},
and $\mathcal{D}_{ik}^{\mu\nu}$ \eqref{eq:oB:Dik},
which describe mutual friction \eqref{eq:general:f} and diffusion \eqref{eq:general:dj} effects,
for the system with two preferred directions.
These coefficients have similar tensor structure
and can be presented as a sum of six symmetric
and three antisymmetric tensor terms,
which are purely spatial in the comoving frame,
and describe anisotropy of mutual friction and diffusion effects in such a system.
The Onsager principle \eqref{eq:oB:Aik=Aki}--\eqref{eq:oB:Cik=Bki}
reduces the number of independent coefficients,
imposing additional
constraints
on $\mathcal{A}_{ik}^{\mu\nu}$
and $\mathcal{D}_{ik}^{\mu\nu}$,
and allowing to express the coefficients 
$\mathcal{C}_{ik}^{\mu\nu}$ through $\mathcal{B}_{ik}^{\mu\nu}$.
Note also that the transport coefficients
(and, consequently, the quantities $f_{(i)}^\mu$ and $\Delta j_{(i)}^\mu$)
have both dissipative and nondissipative 
contributions,
i.e., not all the terms in the expressions for $f_{(i)}^\mu$ and $\Delta j_{(i)}^\mu$
lead to entropy generation in Eq.\ \eqref{eq:dSmu-diff-mf}.

\subsection{One preferred direction}
Now let us assume that there is only one preferred direction in the system, $b^\mu = \omega^\mu$,
i.e., either proton and neutron vortices are aligned with each other,
or there is only one sort of vortices in the system.
In this case, the expressions~\eqref{eq:oB:Aik-2} and \eqref{eq:oB:Bik}--\eqref{eq:oB:Dik}
acquire
the same form as the diffusion coefficients from DGS20,
\begin{gather}
\label{eq:oB:Aik-3}
	\mathcal{A}_{ik}^{\mu\nu}
	= \mathcal{A}_{ik}^{\parallel} \omega^\mu \omega^\nu
	+ \mathcal{A}_{ik}^{\perp} \left( \perp^{\mu\nu} - \omega^\mu \omega^\nu \right)
	+ \mathcal{A}_{ik}^{H} \omega^{\mu\nu}
,\\
\label{eq:oB:Bik-3}
	\mathcal{B}_{ik}^{\mu\nu}
	= \mathcal{B}_{ik}^{\parallel} \omega^\mu \omega^\nu
	+ \mathcal{B}_{ik}^{\perp} \left( \perp^{\mu\nu} - \omega^\mu \omega^\nu \right)
	+ \mathcal{B}_{ik}^{H} \omega^{\mu\nu}
,\\
\label{eq:oB:Cik-3}
	\mathcal{C}_{ik}^{\mu\nu}
	= \mathcal{C}_{ik}^{\parallel} \omega^\mu \omega^\nu
	+ \mathcal{C}_{ik}^{\perp} \left( \perp^{\mu\nu} - \omega^\mu \omega^\nu \right)
	+ \mathcal{C}_{ik}^{H} \omega^{\mu\nu}
,\\
\label{eq:oB:Dik-3}
	\mathcal{D}_{ik}^{\mu\nu}
	= \mathcal{D}_{ik}^{\parallel} \omega^\mu \omega^\nu
	+ \mathcal{D}_{ik}^{\perp} \left( \perp^{\mu\nu} - \omega^\mu \omega^\nu \right)
	+ \mathcal{D}_{ik}^{H} \omega^{\mu\nu}
,
\end{gather}
where
$\mathcal{A}_{ik}^{\parallel} \equiv
\mathcal{A}_{ik}^{\perp} + \mathcal{A}_{ik}^{\omega\omega} + \mathcal{A}_{ik}^{bb} + 2 \mathcal{A}_{ik}^{\omega b}$,
$\mathcal{A}_{ik}^{H} \equiv \mathcal{A}_{ik}^{\omega} + \mathcal{A}_{ik}^{b}$,
and analogous definitions apply to $\mathcal{B}_{ik}^{\parallel}$, $\mathcal{B}_{ik}^{H}$,
$\mathcal{C}_{ik}^{\parallel}$, $\mathcal{C}_{ik}^{H}$,
$\mathcal{D}_{ik}^{\parallel}$, and $\mathcal{D}_{ik}^{H}$.
The Onsager relations \eqref{eq:oB:Aik=Aki-2}, \eqref{eq:oB:Cik=Bki-2}, and \eqref{eq:oB:Dik=Dki-2} then imply
\begin{gather}
\label{eq:oB:single:Aik=Aki}
	\mathcal{A}_{ik}^\parallel = \mathcal{A}_{ki}^\parallel
,\quad
	\mathcal{A}_{ik}^\perp = \mathcal{A}_{ki}^\perp
,\quad
	\mathcal{A}_{ik}^H = \mathcal{A}_{ki}^H
,\\
\label{eq:oB:single:Bik=Cki}
	\mathcal{C}_{ik}^\parallel = - \mathcal{B}_{ki}^\parallel
,\quad
	\mathcal{C}_{ik}^\perp = - \mathcal{B}_{ki}^\perp
,\quad
	\mathcal{C}_{ik}^H = - \mathcal{B}_{ki}^H
,\\
\label{eq:oB:single:Dik=Dki}
	\mathcal{D}_{ik}^\parallel = \mathcal{D}_{ki}^\parallel
,\quad
	\mathcal{D}_{ik}^\perp = \mathcal{D}_{ki}^\perp
,\quad
	\mathcal{D}_{ik}^H = \mathcal{D}_{ki}^H
.
\end{gather}
The coefficients
$\mathcal{A}_{ki}^H$, $\mathcal{D}_{ki}^H$,
$\mathcal{B}_{ik}^\parallel$,
$\mathcal{B}_{ik}^\perp$,
$\mathcal{C}_{ik}^\parallel$,
and $\mathcal{C}_{ik}^\perp$
are nondissipative.

\subsection{Summary}

To sum up, in this section we found a general form of
the four-vectors $f_{(i)}^\mu$ \eqref{eq:general:f},
which encode
all the information about the
forces acting on neutron and proton vortices,
and the diffusive currents $\Delta j_{(i)}^\mu$ \eqref{eq:general:dj},
which describe diffusion, thermodiffusion and thermal conductivity effects.
These vectors are expressed as linear combinations of the vectors $W_{(k)\nu}$ and $d_{(k)\nu}$.
The transport coefficients
$\mathcal{A}_{ik}^{\mu\nu}$, $\mathcal{B}_{ik}^{\mu\nu}$, $\mathcal{C}_{ik}^{\mu\nu}$, and $\mathcal{D}_{ik}^{\mu\nu}$
in these relations
depend on the directions of neutron vortices and the magnetic field;
they are
given by Eqs.~\eqref{eq:oB:Aik-2} and \eqref{eq:oB:Bik}--\eqref{eq:oB:Dik},
which reduce to Eqs.~\eqref{eq:oB:Aik-3}--\eqref{eq:oB:Dik-3} in the case of single preferred direction.
The transport coefficients satisfy the Onsager relations \eqref{eq:oB:Aik=Aki}--\eqref{eq:oB:Cik=Bki}, which 
imply Eqs.~\eqref{eq:oB:Aik=Aki-2}, \eqref{eq:oB:Cik=Bki-2}, and \eqref{eq:oB:Dik=Dki-2}
for a system with two preferred directions,
and Eqs.~\eqref{eq:oB:single:Aik=Aki}--\eqref{eq:oB:single:Dik=Dki} for a system with a single preferred direction.

We emphasize the presence of
cross-coefficients $\mathcal{B}_{ik}^{\mu\nu}$ and $\mathcal{C}_{ik}^{\mu\nu}$,
describing the interplay
of diffusion and mutual friction effects:
the diffusive forces $d_{(k)\nu}$ affect particle velocities (or currents $\Delta j_{(i)}^\mu$),
which, in turn,
influence
the vortex motion via the mutual friction mechanism (and vice versa).

\section{Diffusion and mutual friction in NS matter: special cases}
\label{sec:special-cases}

Let us apply the general formulas from the previous section to a number of
interesting limiting cases, in which these formulas can be substantially simplified.

\subsection{Isotropic matter: neutrons are superfluid, protons are superconducting, no vortices}

In the absence of vortices and any preferred direction
the four-vectors $f_{(i)}^\mu$ vanish in view of Eqs.\ \eqref{eq:Vmunu=0} and \eqref{eq:sfl-eqn}.
Therefore, due to Eqs.~\eqref{eq:general:f} and \eqref{eq:oB:Cik=Bki},
$\mathcal{A}_{ik}^{\mu\nu} = \mathcal{B}_{ik}^{\mu\nu} = \mathcal{C}_{ik}^{\mu\nu}= 0$.
As in normal (nonsuperfluid and nonsuperconducting) MHD (see DGS20), the generalized diffusion coefficient
$\mathcal{D}_{ik}^{\mu\nu}$
in the isotropic matter is then simply given by
\begin{gather}
\label{eq:Dik-isotropic}
	\mathcal{D}_{ik}^{\mu\nu} 
	= \mathcal{D}_{ki}^{\mu\nu}
	= \mathcal{D}_{ik} \perp^{\mu\nu}
,
\end{gather}
and the entropy generation equation \eqref{eq:dSmu-diff-mf}
reduces to
\begin{gather}
\partial_{\mu}
	\left(
		S u^\mu	
		- \frac{\mu_i}{T} \Delta j_{(i)}^\mu
	\right)
=
	\mathcal{D}_{ik} d_{(i)\mu} d_{(k)\mu} 
.
\end{gather}
The generalized diffusion coefficients $\mathcal{D}_{ik}$
in superfluid matter
can be expressed through the momentum transfer rates of microscopic theory similarly to how it is done in
DGS20 for normal matter \cite{gg21}.

\subsection{Magnetized $npe\mu$ matter with superfluid neutrons and normal protons, no vortices}

Now let us consider magnetized $npe\mu$ matter
with superfluid neutrons in the absence of vortices.
Then the only preferred direction is that of the magnetic field, $b^\mu$.
The four-vector $f_{(i)}^\mu$ vanishes in view of Eqs. \eqref{eq:Vmunu=0} and \eqref{eq:sfl-eqn},
but $W_{(i)}^\mu$, generally,
differs from zero.
Therefore, due to Eqs.~\eqref{eq:general:f} and \eqref{eq:oB:single:Bik=Cki},
$\mathcal{A}_{ik}^{\mu\nu} = \mathcal{B}_{ik}^{\mu\nu} = \mathcal{C}_{ik}^{\mu\nu}= 0$.
As a result, $\Delta j_{(i)}^\mu$ has exactly the same form as in the nonsuperfluid magnetized matter (cf.\ DGS20):
\begin{gather}
\label{eq:sfl-n-noV:dj}
\Delta j_{(i)}^\mu
= 
- \mathcal{D}_{ik}^\parallel b^\mu b^\nu  d_{(k)\nu}
- \mathcal{D}_{ik}^\perp \left( \perp^{\mu\nu} -b^\mu b^\nu \right)  d_{(k)\nu}
- \mathcal{D}_{ik}^H b^{\mu\nu}  d_{(k)\nu}
,
\end{gather}
where $i,k=n,p,e,\mu$.
The entropy generation equation \eqref{eq:dSmu-diff-mf}
reduces to
\begin{gather}
\partial_{\mu} 
	\left(
		S u^\mu	
		- \frac{\mu_i}{T} \Delta j_{(i)}^\mu
	\right)
=
	\mathcal{D}_{ik}^\parallel
		b^\mu b^\nu
		d_{(i)\mu} d_{(k)\nu} 
	+ \mathcal{D}_{ik}^\perp
		\left( \perp^{\mu\nu} -b^\mu b^\nu \right)
		d_{(i)\mu} d_{(k)\nu} 
.
\end{gather}
%

\subsection{Unmagnetized $npe\mu$ matter with superfluid neutron vortices}
\label{sec:special-cases:sfl-nV}

In this example, we discuss
the unmagnetized $npe\mu$ matter,
allowing for the presence of superfluid neutron vortices and diffusion.
Protons can be either normal or superconducting.
The dynamic equations for such system
allow us
to simultaneously study
the combined effect of particle diffusion \cite{kgk21} and mutual friction dissipation \cite{haskell15}
on damping of NS oscillations and development of various instabilities in NSs.

Since in real NSs
the typical areal density of neutron vortices is small \cite{gas11}
(the intervortex spacing is much larger than the particle mean free path),
they have a negligible effect on the diffusion coefficients $\mathcal{D}_{ik}^{\mu\nu}$,
which remain approximately isotropic.
Because of the same reason,
the difference between
the velocities of normal particle species
(e.g., electrons and muons or electrons and neutron Bogoliubov thermal excitations)
is small in comparison to the difference between 
any of these velocities
and the neutron vortex velocity, ${\pmb V}_{{\rm L} n}$. 
Consequently, when calculating the force acting on neutron vortices from a particle species $i$
[see Eq.~\eqref{eq:mf:Fj}, where a similar force on proton vortices is presented],
one can replace
${\pmb V}_{i} - {\pmb V}_{{\rm L}n}$
with 
${\pmb V}_{\rm norm} - {\pmb V}_{{\rm L}n}$,
where ${\pmb V}_{\rm norm}$
is the average velocity of normal (nonsuperfluid)
component
\eqref{eq:nonrel:u}.
This approximation allows one to neglect
the cross-coefficients $\mathcal{B}_{ik}^{\mu\nu}$ and $\mathcal{C}_{ik}^{\mu\nu}$,%
\footnote{In principle, these coefficients can be calculated in exactly the same way as it is done
	for superfluid and superconducting $npe\mu$-matter with proton flux tubes in Appendix~\ref{sec:mf} (see also Sec.\ \ref{sec:special-cases:sfl-p}).
	Note, however, that the typical areal density of proton flux tubes in NSs is
	comparable to particle mean free path \cite{gusakov19}, hence
	the cross-coefficients $\mathcal{B}_{ik}^{\mu\nu}$ and $\mathcal{C}_{ik}^{\mu\nu}$
	for this problem
	are not small and should be accounted for.}
%
that is, to decouple
the diffusion and mutual friction mechanisms.
As a result, with the help of Eqs.~\eqref{eq:oB:Aik-3} and \eqref{eq:Dik-isotropic},
Eqs.~\eqref{eq:general:f}--\eqref{eq:general:dj}
reduce to
\begin{gather}
\label{eq:sfl-nV:fn}
- \frac{\mu_n n_n^2}{T} f_{(n)}^\mu
	= - \mathcal{A}_{nn}^\parallel \omega^\mu \omega^\nu  W_{(n)\nu}
	- \mathcal{A}_{nn}^\perp \left( \perp^{\mu\nu} -\omega^\mu \omega^\nu \right)  W_{(n)\nu}
	- \mathcal{A}_{nn}^H \omega^{\mu\nu}  W_{(n)\nu}
,\\
\label{eq:sfl-nV:fp}
	f_{(p)}^\mu = 0
,\\
\label{eq:sfl-nV:dj}
	\Delta j_{(i)}^\mu
	= 
	- \mathcal{D}_{ik} d_{(k)}^\mu
.
\end{gather}
Here the coefficients $\mathcal{A}_{nn}^\perp$, $\mathcal{A}_{nn}^\parallel$, and $\mathcal{A}_{nn}^H$
describe the mutual friction effect.
In order to relate them to the commonly used mutual friction 
parameters
$\alpha_n$, $\beta_n$, and $\gamma_n$ \cite{khalatnikov00,gusakov16,gd16},
one has to compare Eq.~\eqref{eq:sfl-nV:fn} with the analogous equation (98) in GD16, which reads, in our notation,
\begin{gather}
\label{eq:fn-gd16}
	f_{(n)}^\mu
	= \alpha_n \mathcal{V}_{({\rm M}n)} \omega^{\mu\nu} W_{(n)\nu}
	+ (\beta_n - \gamma_n) \mathcal{V}_{({\rm M}n)} \omega^{\mu\alpha} \omega^{\nu}_{~\alpha} W_{(n)\nu}
	+ \gamma_n \mathcal{V}_{({\rm M}n)} \perp^{\mu\nu} W_{(n)\nu}
,
\end{gather}
where $\mathcal{V}_{({\rm M}n)}$ is defined by Eq.~\eqref{eq:Vmagn}.
Using the identity $\omega^{\mu\alpha} \omega^{\nu}_{~\alpha} \equiv \perp^{\mu\nu} -\omega^\mu \omega^\nu$,
we find
\begin{gather}
\label{eq:Ann}
	\mathcal{A}_{nn}^H = \frac{\mu_n n_n^2}{c^3 T} \mathcal{V}_{({\rm M}n)} \alpha_n
,\quad
	\mathcal{A}_{nn}^\perp = \frac{\mu_n n_n^2}{c^3 T} \mathcal{V}_{({\rm M}n)} \beta_n
,\quad
	\mathcal{A}_{nn}^\parallel = \frac{\mu_n n_n^2}{c^3 T} \mathcal{V}_{({\rm M}n)} \gamma_n
,
\end{gather}
where we, for practical convenience, restored the speed of light $c$.
We should stress that, generally,
diffusion affects the coefficients
$\mathcal{A}_{ik}^{\mu\nu}$ (see Section \ref{sec:special-cases:sfl-p} and Appendix~\ref{sec:mf}),
and they cannot be always expressed only through the mutual friction 
parameters
$\alpha_i$, $\beta_i$, and $\gamma_i$
of nondiffusive superfluid hydrodynamics.

It is also worth noting that,
if we allow for the presence of the magnetic field
(assuming that protons are nonsuperconducting and thus $f_{(p)}^\mu = 0$),
but neglect its effect on the neutron vortices,
then expression \eqref{eq:sfl-nV:fn} for $f_{(n)}^\mu$  will remain the same,
while the expression for $\Delta j_{(i)}^\mu$ should be replaced with Eq.~\eqref{eq:sfl-n-noV:dj}
to account for
anisotropy of diffusion in the magnetic field.

\subsection{Magnetized $npe\mu$ matter with superfluid neutrons (no vortices) and type-II proton superconductivity}
\label{sec:special-cases:sfl-p}

This limit is interesting if we want to study
magnetothermal evolution
in slowly rotating superconducting neutron stars
with type-II proton superconductivity.
It is expected that in this problem neutron vortices do not play a major role \cite{ga16}
and can be neglected in the first approximation.
At the same time,
the combined effect of diffusion (i.e., relative motions of different particle species)
and mutual friction dissipation
related to the presence of
proton vortices (flux tubes)
appears to be crucial for this problem
\cite{gko20} and should be accounted for.
Note that, for instance,
electron-flux tube interaction is comparable to
(and even stronger than) the electron-muon interaction
(see, e.g., Ref.~\cite{gko20} and Appendix~\ref{sec:mf}).
Thus, in contrast to the previous case,
here we cannot decouple diffusion and mutual friction effects.

Since we ignore neutron vortices,
we are left with only one preferred direction, ${\pmb b}$.
The full system of 
dynamic
equations
in this situation
is provided in Sec.~\ref{sec:mhd-full},
and here we only present the expressions for $f_{(i)}^\mu$ and $\Delta j_{(i)}^\mu$.
In the absence of neutron vortices $f_{(n)}^\mu$ vanishes,
as do the coefficients $\mathcal{A}_{nk}^{\mu\nu} = \mathcal{B}_{nk}^{\mu\nu} = \mathcal{C}_{in}^{\mu\nu}= 0$.
Thus, the general form of the vectors
$f_{(i)}^\mu$ and $\Delta j_{(i)}^\mu$ is ($i,k=n,p,e,\mu$)
\begin{gather}
\label{eq:sfl-p:fn=0}
	f_{(n)}^\mu	= 0
,\\
\label{eq:p:fp-short}
	- \frac{\mu_p n_p^2}{T} f_{(p)}^\mu	= - \mathcal{A}_{pp}^{\mu\nu} W_{(p)\nu} -  \mathcal{B}_{pk}^{\mu\nu} d_{(k)\nu}
,\\
\label{eq:p:dj-short}
	\Delta j_{(i)}^\mu	= - \mathcal{C}_{ip}^{\mu\nu} W_{(p)\nu} - \mathcal{D}_{ik}^{\mu\nu} d_{(k)\nu}
,
\end{gather}
or, using Eqs.~\eqref{eq:oB:Aik-3}, \eqref{eq:oB:Bik-3}--\eqref{eq:oB:Dik-3}, and \eqref{eq:oB:single:Bik=Cki}
(with $\omega^\mu$ replaced by $b^\mu$ and with $\omega^{\mu\nu}$ replaced by $b^{\mu\nu}$ )
\begin{gather}
\label{eq:fp}
\begin{split}
- \frac{\mu_p n_p^2}{T} f_{(p)}^\mu
= &- \mathcal{A}_{pp}^\parallel b^\mu b^\nu  W_{(p)\nu}
- \mathcal{A}_{pp}^\perp \left( \perp^{\mu\nu} -b^\mu b^\nu \right)  W_{(p)\nu}
- \mathcal{A}_{pp}^H b^{\mu\nu}  W_{(p)\nu}
\\&
- \mathcal{B}_{pk}^\parallel b^\mu b^\nu  d_{(k)\nu}
- \mathcal{B}_{pk}^\perp \left( \perp^{\mu\nu} -b^\mu b^\nu \right)  d_{(k)\nu}
- \mathcal{B}_{pk}^H b^{\mu\nu}  d_{(k)\nu}
,
\end{split}
\\
\label{eq:sfl-p:dj2}
\begin{split}
\Delta j_{(i)}^\mu
= &
- \mathcal{C}_{ip}^{\parallel} b^\mu b^\nu  W_{(p)\nu}
- \mathcal{C}_{ip}^{\perp} \left( \perp^{\mu\nu} -b^\mu b^\nu \right)  W_{(p)\nu}
- \mathcal{C}_{ip}^{H} b^{\mu\nu}  W_{(p)\nu}
\\&
- \mathcal{D}_{ik}^\parallel b^\mu b^\nu  d_{(k)\nu}
- \mathcal{D}_{ik}^\perp \left( \perp^{\mu\nu} -b^\mu b^\nu \right)  d_{(k)\nu}
- \mathcal{D}_{ik}^H b^{\mu\nu}  d_{(k)\nu}
.
\end{split}
\end{gather}
The phenomenological coefficients in Eqs.~\eqref{eq:fp} and \eqref{eq:sfl-p:dj2}
can be expressed through microscopic quantities (mutual friction 
parameters
and momentum transfer rates),
as shown in Appendix~\ref{sec:mf}
in the simple case of vanishing entrainment and $T=0$.
The cross-terms in Eq.\ (\ref{eq:sfl-p:dj2}),
containing the coefficients
$\mathcal{B}_{pi}^{\parallel} = - \mathcal{C}_{ip}^{\parallel}$,
$\mathcal{B}_{pi}^{\perp} = - \mathcal{C}_{ip}^{\perp}$,
and $\mathcal{B}_{pi}^{H} = - \mathcal{C}_{ip}^{H}$,
lead to
interference
between the diffusion and mutual friction effects.

Note, in passing, that if the neutron vortices are present,
but do not affect the diffusive currents (see Sec.~\ref{sec:special-cases:sfl-nV})
and do not interact with proton vortices,
then the expressions for
$f_{(p)}^\mu$ \eqref{eq:fp}
and $\Delta j_{(i)}^\mu$ \eqref{eq:sfl-p:dj2}
will remain the same,
whereas $f_{(n)}^\mu$ will be given by Eq.~\eqref{eq:sfl-nV:fn}.

\section{Full system of equations in
	the MHD approximation for $npe\mu$-mixture with proton vortices}
\label{sec:mhd-full}

In this section we formulate the full system of MHD equations
for magnetized $npe\mu$ matter, accounting for neutron superfluidity
as well as type-II proton superconductivity,
and adopting the ``MHD approximation''
from GD16.
The resulting set of equations, presented in Sec.~\ref{sec:full-mhd:eqns}, is suitable for, e.g., studying
the combined quasistationary evolution of the magnetic field
and temperature
in slowly rotating superconducting NSs.
For practical convenience, below in this section we do not set $c=1$.

\subsection{``Magnetohydrodynamic'' approximation}
\label{sec:mhd-limit}

First, let us 
briefly summarize the main consequences of
the ``MHD approximation'' formulated in Sec.~VIII of GD16,
which allows us to substantially simplify the general equations of Sec.~\ref{sec:gen-eqns}.
This approximation is mainly based on the fact that,
under typical NS conditions
(and assuming
type-II
proton superconductivity),
the magnetic induction ${\pmb B}$ is much larger than the fields 
${\pmb E}$, ${\pmb D}$, and ${\pmb H}$
defined in the comoving frame.
For actual calculations, one also has to specify a microscopic model
that allows to express the four-vectors $D^\mu$, $H^\mu$, $\mathcal{W}_{({\rm E}i)}^\mu$ and $\mathcal{W}_{({\rm M}i)}^\mu$
through $E^\mu$, $B^\mu$, $\mathcal{V}_{({\rm E}i)}^\mu$ and $\mathcal{V}_{({\rm M}i)}^\mu$.
For definiteness, below we use the simple model of noninteracting vortices from Appendix G2 of GD16;
note, however, that the MHD approximation can be
formulated for other microscopic models
in a similar way.

As discussed in Ref.~\cite{gas11} and GD16, the magnetic field ${\pmb H}$ is related to the magnetic induction ${\pmb B}$ as\footnote{
Some authors (e.g., \cite{ep77,aw08,rw20}) use a different definition for ${\pmb H}$,
identifying it with the critical field $H_{{\rm c}1}$;
we find that definition less convenient since ${\pmb H}$ defined
that way
does not satisfy the Maxwell equation \eqref{eq:curlH}.
Note, however, that both approaches are, in principle, possible
and the resulting equations are completely equivalent
\cite{rw20}.
}
\begin{gather}
	{\pmb H} = {\pmb B} - {\pmb B}_{{\rm V}n} - {\pmb B}_{{\rm V}p}
,
\end{gather}
where ${\pmb B}_{{\rm V}i}$ is the magnetic induction
associated with neutron ($i=n$) or proton ($i=p$) vortices.
In other words, ${\pmb H}$
coincides with
the London field
generated by NS rotation,
$\left| {\pmb H} \right|
	\sim 2 \times 10^{-2} \left[\Omega/(100~{\rm s}^{-1})\right]~{\rm G}
	\ll \left| {\pmb B} \right| \sim 10^{12}~{\rm G}$,
where $\Omega$ is the NS spin frequency.
This field, as well as ${\pmb B}_{{\rm V}n}$%
\footnote{
	${\pmb B}_{{\rm V}i}$ is proportional to the number of vortices per unit area $N_{{\rm V}i}$;
	for typical NS conditions $N_{{\rm V}n}$ is less than $N_{{\rm V}p}$ by more than ten orders of magnitude
	and thus $\left| {\pmb B}_{{\rm V}n} \right| \ll \left| {\pmb B}_{{\rm V}p} \right|$.
}
is neglected in comparison to ${\pmb B}_{{\rm V}p}$
in the MHD approximation: all the magnetic induction is assumed to be locked to proton vortices,
${\pmb B} \approx {\pmb B}_{{\rm V}p}$.

Similarly, the fields ${\pmb D}$ and ${\pmb E}$ are related as
\begin{gather}
	{\pmb D} = {\pmb E} - {\pmb E}_{{\rm V}n} - {\pmb E}_{{\rm V}p}
.
\end{gather}
Here the electric field ${\pmb E}_{{\rm V}i}$ 
is generated by vortex motion, 
${\pmb E}_{{\rm V}i} = - \left(1/c\right) {\pmb V}_{{\rm L}i} \times {\pmb B}_{{\rm V}i}$,
where ${\pmb V}_{{\rm L}i}$ is the vortex velocity, which is assumed to be nonrelativistic;
the electric induction ${\pmb D}$ is of the order of small gradients of thermodynamic functions,
$\left|{\pmb D}\right| \sim \left|{\pmb \nabla} \mu_i \right|/e_p$.
Both ${\pmb E}$ and ${\pmb D}$ are much smaller than ${\pmb B}$.

Since the vectors ${\pmb D}$ and ${\pmb H}$ are small,
it follows from the second pair of Maxwell equations \eqref{eq:maxwell-2}
that the total free electric current density $J_{({\rm free})}^\mu$
should also be exceptionally small,
much smaller than the individual contributions to $J_{({\rm free})}^\mu$ from
each particle species.
This observation enables us to make further simplification by discarding
Maxwell equations \eqref{eq:maxwell-2}, but instead requiring
that
the free electric current density $J_{({\rm free})}^\mu$
should vanish 
[this approximation is well-known in the literature and is further discussed by us around Eqs.\ (\ref{eq:full-mhd:quasineutrality}) and (\ref{eq:full-mhd:screening})],
\begin{gather}
\label{eq:mhd-limit:Jfree=0}
	J^{\mu}_{\rm (free)}
	= e_i n_i u^\mu
		+ e_i Y_{ik} w_{(k)}^\mu
		+ e_i \Delta j_{(i)}^\mu
	= 0
.
\end{gather}

Now let us turn to the vortex-related vectors
$\mathcal{V}_{({\rm E}i)}^\mu$, $\mathcal{V}_{({\rm M}i)}^\mu$, $\mathcal{W}_{({\rm E}i)}^\mu$, and $\mathcal{W}_{({\rm M}i)}^\mu$
[or, equivalently, to the corresponding tensors
$\Vpar_{(i)}^{\mu\nu}$, $\Vperp_{(i)}^{\mu\nu}$, $\Wpar_{(i)}^{\mu\nu}$, and $\Wperp_{(i)}^{\mu\nu}$,
see Eqs.~\eqref{eq:Apar} and \eqref{eq:Aperp}].
The number of proton vortices is typically larger by more than ten orders of magnitude than the number of neutron vortices (see, e.g., Ref.\ \cite{gas11}).
Consequently, the four-vector $\mathcal{V}_{({\rm M}n)}^\mu$
can be neglected in comparison to $\mathcal{V}_{({\rm M}p)}^\mu$
in the expressions for $d \varepsilon_{\rm add}$
\eqref{eq:de-EM-vortex}
and $\Delta T^{\mu\nu}_{({\rm EM + vortex})}$
\eqref{eq:dT-EM-vortex},
since the lengths of these vectors are proportional to the number of vortices, as follows from Eq.~\eqref{eq:Nvi}.
Note also that in the comoving frame $|\mathcal{V}_{({\rm E}i)}| \sim \left( V_{{\rm L}i}/c \right) |\mathcal{V}_{({\rm M}i)}|$,
thus $\mathcal{V}_{({\rm E}i)}^\mu$ can be neglected in comparison to $\mathcal{V}_{({\rm M}i)}^\mu$,
and, similarly,
$\mathcal{W}_{({\rm E}i)}^\mu$
can be neglected in comparison to
$\mathcal{W}_{({\rm M}i)}^\mu$.

Under the above assumptions,
the four-vector
$\widetilde{\mathcal{V}}_{({\rm M}p)}^\mu
\equiv \frac{1}{2} \, \epsilon^{\mu \nu \alpha \beta} \, u_{\nu} \, \widetilde{\mathcal{V}}_{(i) \alpha \beta}$,
which reduces to $(0, m_p {\rm curl} {\pmb V}_{{\rm s}p})$ in the nonrelativistic limit,
can be neglected in comparison to 
$(e_p/c) B^\mu$.
Thus, the four-vector
$\mathcal{V}_{({\rm M}p)}^\mu
= \widetilde{\mathcal{V}}_{({\rm M}p)}^\mu
+ (e_p/c) B^\mu$
[see Eqs.~\eqref{eq:Amagn} and \eqref{eq:tildeV}],
reduces to
\begin{gather}
\label{eq:mhd-limit:VMp}
	\mathcal{V}_{({\rm M}p)}^\mu = \frac{e_p}{c} B^\mu
,
\end{gather}
which physically means that the magnetic induction is produced by proton vortices.

For a simple microscopic model of noninteracting vortices,
the four-vectors $\mathcal{W}_{({\rm M}i)}^\mu$ are related to $\mathcal{V}_{({\rm M}i)}^\mu$ as
[see equations (124) and (G9)--(G11) in GD16]
\begin{gather}
\label{eq:mhd-limit:WMi}
	\mathcal{W}_{({\rm M}i)}^\mu
	= \frac{\hat{E}_{{\rm V}i}}
		   {\pi \hbar}
		\frac{\mathcal{V}_{({\rm M}i)}^\mu}
		   {\mathcal{V}_{({\rm M}i)}}
,
\end{gather}
where $\mathcal{V}_{({\rm M}i)} \equiv \sqrt{\mathcal{V}_{({\rm M}i)\alpha} \mathcal{V}_{({\rm M}i)}^\alpha}$,
$\hat{E}_{{\rm V}i}$ is the vortex energy per unit length specified below, and no summation over $i$ is assumed.
$\mathcal{W}_{({\rm M}p)}^\mu$ can also be rewritten in terms of
the critical magnetic field $H_{\rm c1}$ \cite{ll80},
\begin{gather}
\label{eq:mhd-limit:WMp}
	\mathcal{W}_{({\rm M}p)}^\mu
	= \frac{c}{4\pi e_p} H_{\rm c1} \frac{B^\mu}{B}
.
\end{gather}
In this formula
$B \equiv (B_\mu B^\mu)^{1/2}$,
and $H_{\rm c1}$ is expressed through
$\hat{E}_{{\rm V}p}$ as
\begin{gather}
\label{eq:Hc1}
	H_{\rm c1} = \frac{4\pi \hat{E}_{{\rm V}p}}{\hat{\phi}_{p0}}
,
\end{gather}
where $\hat{\phi}_{p0} = (\pi \hbar c / e_p)$ is the magnetic flux associated with proton vortex.
The energy $\hat{E}_{{\rm V}i}$ per unit length for neutron and proton vortices
is given by [see equations (E17) and (E18) in GD16]
\begin{gather}
\label{eq:EVn}
	\hat{E}_{{\rm V}n}
	\approx \frac{\pi}{4} \hbar^2 c^2
		\frac{Y_{nn} Y_{pp} - Y_{np}^2}{Y_{pp}}
		\ln \left( \frac{b_n}{\xi_n} \right)
,\\
\label{eq:EVp}
	\hat{E}_{{\rm V}p}
	 \approx \frac{\pi}{4}\hbar^2 c^2 Y_{pp} \ln \left(\frac{\delta_p}{\xi_p} \right)
.
\end{gather}
In Eqs.\ (\ref{eq:EVn}) and (\ref{eq:EVp})
$\xi_i$ is the coherence length for particle species $i$,
$\delta_p$ is the London penetration depth for protons,
and $b_n$ is
some ``external'' radius of the order
of the typical intervortex spacing \cite{khalatnikov00,gusakov16}.
Note that Eq.~\eqref{eq:EVp} (see also Ref.~\cite{mendell91a} for a nonrelativistic expression)
is 
only applicable to
a strong type-II superconductor, i.e., in the limit
$\delta_p \gg \xi_p$.

We remind the reader that the expressions \eqref{eq:mhd-limit:WMi} for $\mathcal{W}_{({\rm M}i)}^\mu$ 
are valid only for a simple model of noninteracting vortices.
If one accounts, e.g., for vortex-flux tube interaction, then
both these vectors will depend on $\mathcal{V}_{({\rm M}p)}^\mu$ and $\mathcal{V}_{({\rm M}n)}^\mu$ simultaneously.

Using the approximations discussed above, one can also simplify the thermodynamic relations.
First, all the thermodynamic quantities (e.g., the energy density $\varepsilon$)
can be expressed as functions of the variables
$n_i$, $S$, $w_{(i)}^\mu w_{(k)\mu}$, and $B$,
\begin{gather}
\label{eq:mhd-limit:energy}
	\varepsilon = \varepsilon \left( n_i, S, w_{(i)}^\mu w_{(k)\mu}, B \right)
.
\end{gather}
Second,
only the term $\mathcal{W}_{({\rm M}p)\mu} d \mathcal{V}^{\mu}_{({\rm M}p)}$
can be retained
in the expression \eqref{eq:de-EM-vortex} for 
$d \varepsilon_{\rm add}$.
Thus, in view of the relations \eqref{eq:mhd-limit:VMp}
and \eqref{eq:mhd-limit:WMp},
the second law of thermodynamics \eqref{eq:2ndlaw} 
becomes
\begin{equation}
\label{eq:mhd-limit:2ndlaw}
	d \varepsilon = \mu_i \, dn_i 
		+ T \, dS 
		+ \frac{Y_{ik}}{2} \, d \left( w_{(i)}^\alpha w_{(k)\alpha}  \right)
		+ \frac{1}{4\pi} H_{\rm c1} dB
,
\end{equation}
and the Gibbs-Duhem relation \eqref{eq:dP}, consequently, 
takes the form
\begin{gather}
\label{eq:mhd-limit:dP}
	dP = n_i  \, d\mu_i + S \, dT
	- \frac{Y_{ik}}{2} \, d \left( w_{(i)}^\alpha w_{(k)\alpha}  \right)
	- \frac{1}{4\pi} H_{\rm c1} dB
.
\end{gather}
Similarly, only the last term (and only for proton vortices, $i=p$) survives in the expression for $\Delta T^{\mu\nu}_{({\rm EM + vortex})}$ \eqref{eq:dT-EM-vortex},
\begin{gather}
\label{eq:mhd-limit:dT-EM-vortex}
	\Delta T^{\mu\nu}_{({\rm EM + vortex})}
	= \mathcal{T}^{\mu\nu}_{({\rm VM})}
	= 	\Wperp^{\mu\alpha}_{(p)}\Vperp^{\nu}_{(p)\,\alpha}
		+ u^{\nu} \Wperp^{\mu\alpha}_{(p)} \mathcal{V}_{({\rm E}p)\alpha}
		+u^{\mu}\Wperp^{\nu\alpha}_{(p)}\mathcal{V}_{({\rm E}p)\alpha}
.
\end{gather}
Noting that
$\mathcal{V}_{({\rm E}p)}^{\mu} = \mu_p n_p f_{(p)}^\mu / c^3$
[see Eq.~\eqref{eq:sfl-eqn}],
and also using the relations \eqref{eq:mhd-limit:VMp} and \eqref{eq:mhd-limit:WMp},
one can transform Eq.~\eqref{eq:mhd-limit:dT-EM-vortex} to
\begin{gather}
\label{eq:full-mhd:TVM-1}
	\Delta T^{\mu\nu}_{({\rm EM + vortex})}
	=
	\frac{H_{\rm c1} B}{4 \pi} b^{\mu\alpha} b^{\nu}_{\phantom{\nu}\alpha} 
	+ \frac{\mu_p n_p H_{\rm c1}}{4\pi e_p c^2}
		\left(
			u^\mu b^{\nu\alpha} f_{(p)\alpha}
		  + u^\nu b^{\mu\alpha} f_{(p)\alpha}
		\right)
,
\end{gather}
or, equivalently, to
\begin{gather}
\label{eq:full-mhd:TVM-2}
	\Delta T^{\mu\nu}_{({\rm EM + vortex})}
	=
	\frac{H_{\rm c1} B}{4 \pi} \left( \perp^{\mu\nu} - b^\mu b^\nu \right) 
	+ \frac{\mu_p n_p H_{\rm c1}}{4\pi e_p c^2}
		\left(
			u^\mu \epsilon^{\nu\alpha\beta\gamma} u_\alpha f_{(p)\beta} b_\gamma 
		  + u^\nu \epsilon^{\mu\alpha\beta\gamma} u_\alpha f_{(p)\beta} b_\gamma 
		\right)
.	
\end{gather}
Repeating the derivation of the entropy generation equation \eqref{eq:dSmu-EM2}
with $d \varepsilon$ given by Eq.~\eqref{eq:mhd-limit:2ndlaw}
and $\Delta T^{\mu\nu}_{({\rm EM + vortex})}$ given by Eq.~\eqref{eq:mhd-limit:dT-EM-vortex},
one can find that the four-vectors $W_{(n)}^\mu$ and $W_{(p)}^\mu$
[see Eq.~\eqref{eq:Wmu}]
in the MHD limit
should be defined as
\begin{gather}
\label{eq:mhd-limit:Wn}
	W^{\mu}_{(n)}
	\equiv \frac{1}{n_n} c Y_{nk} w^\mu_{(k)} 
,\\
\label{eq:mhd-limit:Wp}
W^{\mu}_{(p)}
	= 
	\frac{1}{n_p}\left[
	c Y_{pk} w^\mu_{(k)}
	+ \frac{c}{4\pi e_p} 
		\perp^{\mu\nu} \partial^{\alpha} \left(
		H_{\rm c1} b_{\nu\alpha}
	\right)	\right]	
.
\end{gather}

\subsection{MHD equations}
\label{sec:full-mhd:eqns}

Now, working in the MHD approximation 
described above, let us formulate the 
dynamic
equations for superconducting NSs
with $npe\mu$ cores.
We assume that protons form a type-II superconductor,
and neutrons are superfluid.
However, we ignore the effects of NS rotation
and hence assume that there are no neutron vortices in the system,
$\mathcal{V}_{(n)}^{\mu\nu} = 0$.
Note that neutron vortices can be included separately (see Remark 3).
As for the dissipative effects, we consider only diffusion and mutual friction,
thus ignoring chemical reactions as well as viscosity
(i.e., we set $Q = \Delta \Gamma_i = \Delta \tau^{\mu\nu} = \varkappa_i = 0$).
The latter effects can easily be incorporated separately if needed.

The full set of equations 
allows to find
seven unknown functions
$B^\mu$, $u^\mu$, $w_{(n)}^\mu$, $n_n$, $n_e$, $n_\mu$, and $S$
(all other unknown quantities can be expressed algebraically through these functions)
and includes:
\begin{enumerate}
\item 
Continuity equations for neutrons, electrons, and muons 
describing
evolution of $n_n$, $n_e$, and $n_\mu$, respectively:
\begin{gather}
\label{eq:full-mhd:jn}
	\partial_\alpha j_{(n)}^\alpha
	= \partial_\alpha \left( n_n u^\alpha + Y_{nk} w_{(k)}^\alpha + \Delta j_{(n)}^\alpha \right)
	= 0
,\\
\label{eq:full-mhd:je}
	\partial_\alpha j_{(e)}^\alpha
	= \partial_\alpha \left( n_e u^\alpha + \Delta j_{(e)}^\alpha \right)
	= 0
,\\
\label{eq:full-mhd:jmu}
	\partial_\alpha j_{(\mu)}^\alpha
	= \partial_\alpha \left( n_{\mu} u^\alpha + \Delta j_{(\mu)}^\alpha \right)
	= 0
.
\end{gather}

\item
Total energy ($\mu=0$) and momentum ($\mu=1,2,3$) conservation laws \eqref{eq:dTmunu=0}
describing
evolution of the energy density $\varepsilon$ and four-velocity $u^\mu$:
\begin{gather}
\partial_{\nu} T^{\mu\nu} = 0 
,
\end{gather}
where
\begin{gather}
\label{eq:full-mhd:Tmunu}
	T^{\mu\nu}
	= {(P+\varepsilon) u^{\mu} u^\nu}
		+ {P g^{\mu\nu}} 
		+ Y_{ik} \left(
			w_{(i)}^\mu w_{(k)}^\nu
			+ \mu_i w_{(k)}^\mu u^\nu
			+ \mu_k w_{(i)}^\nu u^\mu
		\right)
		+ \Delta T^{\mu\nu}_{({\rm EM + vortex})}
,
\end{gather}
and $\Delta T^{\mu\nu}_{({\rm EM+vortex})}$ is specified by Eq.~\eqref{eq:full-mhd:TVM-1}.
Instead of the energy conservation law, it is convenient to use
the entropy generation equation \eqref{eq:dSmu-diff-mf},
\begin{gather}
\label{eq:full-mhd:dS}
\partial_{\mu} S^\mu =
	\partial_{\mu} \left(
		S u^\mu
		- \frac{\mu_i}{T} \Delta j_{(i)}^\mu
		\right)
= 	\frac{\mu_p n_p^2}{c^3 T}
f_{(p)\mu} W^{\mu}_{(p)}
- \Delta j_{(i)}^\mu d_{(i)\mu}
.
\end{gather}

\item
The four-vector $w^\mu_{(n)}$ satisfies the superfluid equation for neutrons, which,
in the absence of vortices, reads
\begin{gather}
\label{eq:full-mhd:sfl-n}
\mathcal{V}^{\mu\nu}_{(n)}
\equiv
\frac{1}{c}
\left[
	\partial^\mu\left( w^\nu_{(n)}+\mu_n u^{\nu} \right)
	-\partial^\nu\left(w^\mu_{(n)}+\mu_n u^{\mu} \right)
\right]
= 0
.
\end{gather}

\item 
Magnetic induction evolves according to Maxwell equation \eqref{eq:maxwell-1},
\begin{gather}
\label{eq:full-mhd:maxwell-1}
	\partial_{\mu} F_{\nu\lambda}
	+ \partial_{\nu} F_{\lambda\mu}
	+ \partial_{\lambda} F_{\mu\nu}
	= 0
,
\end{gather}
which, in terms of the vectors ${\pmb E}$ and ${\pmb B}$,
reads
\begin{gather}
\label{eq:full-mhd:curl-E}
	{\rm curl} {\pmb E}
	= - \frac{1}{c} \pd{\pmb B}{t}
,\\
\label{eq:full-mhd:divB}
	{\rm div} {\pmb B} = 0
.
\end{gather}
\end{enumerate}

The set of equations
\eqref{eq:full-mhd:jn}--\eqref{eq:full-mhd:divB}
contains also unknown quantities $n_p$, $w_{(p)}^\mu$, ${\pmb E}$, $f_{(p)}^\mu$, and $\Delta j_{(p)}^\mu$,
which are expressed algebraically through
the seven functions defined above.

First, the quantities $n_p$ and $w_{(p)}^\mu$
can be found from 
the condition $J_{({\rm free})}^\mu = 0$ \eqref{eq:mhd-limit:Jfree=0},
which, in view of the constraints \eqref{eq:uw=0} and \eqref{eq:udj=0},
leads to the well-known (and often employed in the literature)
quasineutrality \eqref{eq:full-mhd:quasineutrality} and screening \eqref{eq:full-mhd:screening} conditions \cite{jones91, gas11, gd16}:
\begin{gather}
\label{eq:full-mhd:quasineutrality}
	n_p = n_e + n_\mu
,\\
\label{eq:full-mhd:screening}
	Y_{pk} w_{(k)}^\mu + \Delta j_{(p)}^\mu - \Delta j_{(e)}^\mu - \Delta j_{(\mu)}^\mu = 0
.
\end{gather}
Next,
the quantities $f_{(p)}^\mu$ \eqref{eq:fp} and $\Delta j_{(i)}^\mu$ \eqref{eq:sfl-p:dj2} have the following form
[note that we restored the factor $c^3$ in the left-hand side of Eq.~\eqref{eq:full-mhd:fp}]:
\begin{gather}
\label{eq:full-mhd:fp}
\begin{split}
- \frac{\mu_p n_p^2}{c^3 T} f_{(p)}^\mu
= &- \mathcal{A}_{pp}^\parallel b^\mu b^\nu  W_{(p)\nu}
- \mathcal{A}_{pp}^\perp \left( \perp^{\mu\nu} -b^\mu b^\nu \right)  W_{(p)\nu}
- \mathcal{A}_{pp}^H b^{\mu\nu}  W_{(p)\nu}
\\&
- \mathcal{B}_{pk}^\parallel b^\mu b^\nu  d_{(k)\nu}
- \mathcal{B}_{pk}^\perp \left( \perp^{\mu\nu} -b^\mu b^\nu \right)  d_{(k)\nu}
- \mathcal{B}_{pk}^H b^{\mu\nu}  d_{(k)\nu}
,
\end{split}
\\
\label{eq:full-mhd:dj}
\begin{split}
\Delta j_{(i)}^\mu
= &-\mathcal{C}_{ip}^{\parallel} b^\mu b^\nu  W_{(p)\nu}
- \mathcal{C}_{ip}^{\perp} \left( \perp^{\mu\nu} -b^\mu b^\nu \right)  W_{(p)\nu}
- \mathcal{C}_{ip}^{H} b^{\mu\nu}  W_{(p)\nu}
\\&
- \mathcal{D}_{ik}^\parallel b^\mu b^\nu  d_{(k)\nu}
- \mathcal{D}_{ik}^\perp \left( \perp^{\mu\nu} -b^\mu b^\nu \right)  d_{(k)\nu}
- \mathcal{D}_{ik}^H b^{\mu\nu}  d_{(k)\nu}
,
\end{split}
\end{gather}
where $d_{(i)\mu}$ and $W_{(p)\nu}$
are given by Eqs.~\eqref{eq:dmu}
and \eqref{eq:mhd-limit:Wp},
respectively.
The transport coefficients
$\mathcal{A}_{pp}^\parallel$, $\mathcal{A}_{pp}^\perp$, $\mathcal{A}_{pp}^H$,
$\mathcal{B}_{pk}^\parallel$, $\mathcal{B}_{pk}^\perp$,  $\mathcal{B}_{pk}^H$,
$\mathcal{C}_{ip}^\parallel$, $\mathcal{C}_{ip}^\perp$,  $\mathcal{C}_{ip}^H$,
$\mathcal{D}_{ik}^\parallel$, $\mathcal{D}_{ik}^\perp$, and $\mathcal{D}_{ik}^H$
should be expressed through microscopic mutual friction 
parameters
and momentum transfer rates.
We discuss these relations in Appendix~\ref{sec:mf}.

Finally, the electric field $E^\mu$ can be expressed algebraically from the superfluid proton equation \eqref{eq:sfl-eqn},
\begin{gather}
\label{eq:full-mhd:sfl-p}
	u_{\nu} \mathcal{V}_{(p)}^{\mu\nu}
	\equiv \frac{1}{c} u_\nu
		\left\{
		\partial^\mu\left[ w^\nu_{(p)}+\mu_p u^{\nu}\right]
		-\partial^\nu\left[ w^\mu_{(p)}+\mu_p u^{\mu} \right]
		\right\}
		+ \frac{e_p}{c} E^\mu
	= \frac{\mu_p n_p}{c^3} f_{(p)}^\mu
.
\end{gather}
Note that the right-hand sides of Eqs.~\eqref{eq:full-mhd:fp} and \eqref{eq:full-mhd:dj}
implicitly contain $\Delta j_{(i)}^\mu$ 
and $E^\mu$,\footnote{%
\label{footnote:implicitly-contains}
	$W_{(p)}^\mu$
	depends on
	the quantity $Y_{pk} w_{(k)}^\mu$ [see the definition \eqref{eq:Wmu}],
	which is expressed through	$\Delta j_{(i)}^\mu$ 
	with the help of the screening condition \eqref{eq:full-mhd:screening}.
	In addition, $d_{(k)}^\mu$ 
	depends on
	$E^\mu$ [see the definition \eqref{eq:dmu}].
}
therefore one has to solve equations
\eqref{eq:full-mhd:fp}, \eqref{eq:full-mhd:dj}, and \eqref{eq:full-mhd:sfl-p}
simultaneously in order to obtain
closed-form
expressions for $f_{(p)}^\mu$, $\Delta j_{(i)}^\mu$, and $E^\mu$.

The nonrelativistic version of MHD equations from this section is provided
in Appendix~\ref{sec:mhd-full-nonrel}.

\vspace{0.2 cm}
\noindent
%
{\bf Remark 1.}
If $\mathcal{A}_{pp}^\parallel = \mathcal{B}_{pi}^{\parallel} = 0$,
one can define the
vortex velocity $v_{({\rm L}p)}^\mu$,
satisfying the vorticity transfer equation
\cite{gusakov16}
\begin{gather}
v_{({\rm L}p)\nu}  \mathcal{V}_{(p)}^{\mu\nu} = 0
.
\end{gather}
In analogy with GD16 [see equation (101) there], one can find that,
up to arbitrary terms parallel to $b^\mu$,
\begin{gather}
\label{eq:full-mhd:vLp}
v_{({\rm L}p)}^\mu
	= u^\mu
	- \frac{cT}{n_p e_p B}
	\left( \mathcal{A}_{pp}^H W_{(p)\nu} + \mathcal{B}_{pk}^H d_{(k)\nu} \right) \perp^{\mu\nu}
	+ \frac{cT}{n_p e_p B}
	\left( \mathcal{A}_{pp}^\perp W_{(p)\nu} + \mathcal{B}_{pk}^\perp d_{(k)\nu} \right) b^{\mu\nu}
.
\end{gather}

\vspace{0.2 cm}
\noindent
%
{\bf Remark 2.}

The MHD equations presented in this section are very similar to those of Sec.~VIII in GD16.
For the reader's convenience, let us list their main differences from GD16:
\begin{enumerate}
\item
Particle currents include the dissipative corrections $\Delta j_{(i)}^\mu$.
\item
We use a slightly different definition of $W_{(p)}^\mu$
(see footnote \ref{footnote:Wmu}).
\item
The term $Y_{pk}w_{(k)}^\mu$ in the expression \eqref{eq:mhd-limit:Wp} for $W_{(p)}^\mu$ 
does not vanish due to the presence of diffusive currents.
\item
$f_{(p)}^\mu$ (and thus $v_{({\rm L}p)}^\mu$) includes additional terms
proportional to $d_{(k)}^\mu$
(if transport coefficients $\mathcal{B}_{ik}^{\mu\nu} \neq 0$). 
\item
Neutron vortices are absent, $\mathcal{V}_{(n)}^{\mu\nu} = 0$.
\end{enumerate}

\vspace{0.2 cm}
\noindent
%
{\bf{Remark 3.}}

One can easily account for the presence of neutron vortices,
provided that we neglect
their effect on diffusion and ignore vortex-flux tube interaction
(see Sec.~\ref{sec:special-cases:sfl-nV}).
Under these assumptions,
all equations of this section remain the same,
except for Eq.~\eqref{eq:full-mhd:sfl-n}, 
which should be replaced with 
\begin{gather}
\label{eq:full-mhd:sfl-n-vortices}
	u_{\nu} \mathcal{V}_{(n)}^{\mu\nu}
	\equiv \frac{1}{c} u_\nu
	\left\{
	\partial^\mu\left[ w^\nu_{(n)}+\mu_n u^{\nu}\right]
	-\partial^\nu\left[ w^\mu_{(n)}+\mu_n u^{\mu} \right]
	\right\}
	= \frac{\mu_n n_n}{c^3} f_{(n)}^\mu
,
\end{gather}
and Eq.~\eqref{eq:full-mhd:dS}, which should be replaced with
\begin{gather}
\partial_{\mu} S^\mu =
	\partial_{\mu} \left(
		S u^\mu
		- \frac{\mu_i}{T} \Delta j_{(i)}^\mu
		\right)
= 	\frac{\mu_p n_p^2}{c^3 T}
        f_{(p)\mu} W^{\mu}_{(p)}
    + \frac{\mu_n n_n^2}{c^3 T}
        f_{(n)\mu} W^{\mu}_{(n)}
- \Delta j_{(i)}^\mu d_{(i)\mu}
,
\end{gather}
where $f_{(n)}^\mu$
is [see Eq.~\eqref{eq:sfl-nV:fn}]
\begin{gather}
\label{eq:full-mhd:fn}
	- \frac{\mu_n n_n^2}{c^3 T} f_{(n)}^\mu
	= - \mathcal{A}_{nn}^\parallel \omega^\mu \omega^\nu  W_{(n)\nu}
		- \mathcal{A}_{nn}^\perp \left( \perp^{\mu\nu} -\omega^\mu \omega^\nu \right)  W_{(n)\nu}
		- \mathcal{A}_{nn}^H \omega^{\mu\nu}  W_{(n)\nu}
,
\end{gather}
and $W_{(n)}^\mu$
is given by Eq.~\eqref{eq:mhd-limit:Wn}.

\section{Summary}
\label{sec:conclusion}

In the present study we have formulated equations of dissipative relativistic finite-temperature MHD
describing superfluid/superconducting charged mixtures in the presence of vortices and electromagnetic field.
For the first time, the corresponding MHD equations systematically and simultaneously take into account
the combined effects of particle diffusion
and mutual friction forces acting on superfluid/superconducting vortices.
It is important to stress that these two effects interfere with one another:
diffusion affects particle velocities
which, in turn, influences the vortex motion via the mutual friction mechanism (and vice versa);
as a result,
the cross-coefficients $\mathcal{B}_{ik}^{\mu\nu}$ and $\mathcal{C}_{ik}^{\mu\nu}$
in Eqs.~\eqref{eq:general:f} and \eqref{eq:general:dj}
differ from zero.

We have obtained the general MHD equations
and derived the entropy generation equation,
following the same phenomenological approach \cite{bk61,ll87} as in our previous papers \cite{gusakov16,gd16,dgs20} (see Secs.~\ref{sec:gen-eqns} and \ref{sec:entropy}).
These equations extend the results of GD16 (which neglects all the dissipative processes except for the mutual friction dissipation)
by accounting for the diffusion, viscosity, chemical reactions, and radiation.
Then, starting from the Onsager principle and the condition of non-negative entropy production rate,
we have derived
in Sec.~\ref{sec:f-dj}
the general expressions
for the mutual friction forces and diffusive currents
adopting the
``MHD approximation'' from GD16 (see Sec.~\ref{sec:mhd-limit}),
that utilizes the fact that in typical NS conditions
the magnetic induction ${\pmb B}$ is much larger than the fields 
${\pmb E}$, ${\pmb D}$, and ${\pmb H}$.
Note that, in this approximation,
mutual friction and diffusion 
(which are the main focus of our study)
appear to be completely decoupled from other dissipative mechanisms, which can be studied separately.

We have applied the formulated MHD to a number of special cases, 
where it can be considerably simplified (some of these cases are interesting because of their application to NSs).
In particular, 
simplifications arising for unmagnetized
NSs are discussed in Sec.~\ref{sec:special-cases:sfl-nV}.
The resulting equations allow one to easily study
the effect of diffusion and mutual friction dissipation 
on damping of stellar oscillations and
various dynamical instabilities in NSs \cite{kgk21,haskell15}.
In turn, Sec.~\ref{sec:special-cases:sfl-p}
provides equations suitable for studying the quasistationary magnetic field 
evolution
in superconducting NS cores \cite{gko20}.
The full system of equations in this limit is presented in Sec.~\ref{sec:mhd-full}
and 
describes 
$npe\mu$ matter with type-II
proton superconductivity,
accounting for an interplay of mutual friction and particle diffusion dissipation.

The MHD equations discussed above 
contain a number of phenomenological transport coefficients,
that have to be 
determined from microphysics.
We have shown (see Appendix~\ref{sec:mf}) how to establish a connection
between our formalism and the microscopic approach,
by expressing the phenomenological coefficients arising in our theory 
through the microscopic mutual friction parameters
$D_i$ and momentum transfer rates $J_{ik}$
in the low-temperature limit.
We emphasize that \textit{all} these phenomenological coefficients,
generally, depend on both $D_i$ and $J_{ik}$ due to interference between the diffusion and mutual friction
mechanisms.

We see two main immediate practical applications of our results.
First, the dissipative MHD equations, presented in this work, allow one to realistically model long-term
magnetothermal
evolution in superconducting NSs,
accounting for the macroscopic particle flows, diffusive currents, mutual friction, finite temperatures,  as well as special and general relativistic effects.
Second, with the help of these equations, 
one can study the combined effect of diffusion and mutual friction on oscillations and hydrodynamic instabilities in NSs: 
these effects 
are extremely efficient dissipative agents in superfluid and superconducting NS cores \cite{kgk21,haskell15}.

The presented magnetohydrodynamics can be generalized in 
a number of ways.
First, one can easily  consider a more complex particle composition (e.g, including hyperons) within the presented framework.
Another straightforward step is to consider viscosity and chemical reactions 
in the presence of two preferred directions
in the system (specified by the two types of vortices), and to 
derive general form 
of the corresponding dissipative corrections following the same procedure as in Sec.~\ref{sec:f-dj}.
Further, an important task would be
to describe pinning of neutron vortices to proton flux tubes and the vortex creep.
In principle, our general equations should account for these effects,
but for practical applications
one also has to find a relation between 
the phenomenological quantities
(such as the vector $\mathcal{W}_{({\rm M}i)}^\mu$ 
or the transport coefficient $\mathcal{A}^{\mu\nu}_{ik}$) 
and the 
microscopic parameters describing vortex-fluxtube interaction
\cite{als84,sa09,link14,ga16,alpar17,sc20}.
We expect that all these improvements will enable further progress towards realistic modelling of the various dynamical processes in NSs.

\section*{Acknowledgments}

We are very grateful to E.M.~Kantor for reading the draft version of the paper
and valuable comments.
This study is partially supported
by the Foundation for the Advancement of Theoretical Physics and Mathematics BASIS [Grant No. 17-12-204-1]
and by Russian Foundation for Basic Research [Grant No. 19-52-12013].

\appendix

\section{Nonrelativistic limit of equations of Sec.~\ref{sec:mhd-full}}
\label{sec:mhd-full-nonrel}

In this Appendix we present three-dimensional version of MHD equations of Sec.~\ref{sec:mhd-full}
(see analogous equations in Appendix I of GD16), assuming that all macroscopic velocities are nonrelativistic
(the `low-velocity' limit).
At the same time, we employ relativistic equation of state and discuss transition to the fully nonrelativistic limit separately.
To proceed to 
the latter limit,
one has to assume that not only macroscopic velocities, but also equation of state is nonrelativistic.
Then one has to replace the chemical potential $\mu_i$
for particle species $i$ with the particle rest energy, $m_i c^2$
[note, however, that in the superfluid equations for neutrons \eqref{eq:nonrel:sfl-n}
and protons \eqref{eq:nonrel:sfl-p},
as well as in Eq.~\eqref{eq:nonrel:d}
one should retain the small quantity $\breve{\mu}_i \equiv (\mu_i - m_i c^2)/m_i$],
and express the entrainment matrix $Y_{ik}$
through the nonrelativistic matrix $\rho_{ik}$
by the formula \cite{ga06}
\begin{gather}
\label{eq:nonrel:rho-ik}
	\rho_{ik} = m_i m_k c^2 Y_{ik}
,
\end{gather}
where no summation over repeated indices is assumed.
In the absence of entrainment $\rho_{ik} = \rho_{{\rm s}i} \delta_{ik}$,
i.e., the off-diagonal elements of the matrix vanish,
and diagonal elements contain superfluid mass densities $\rho_{{\rm s}i}$ for particle species $i$.
In the fully nonrelativistic limit, the pressure $P$
can be neglected in comparison to the energy density $\varepsilon$,
which equals the rest energy density:
\begin{gather}
\label{eq:nonrel:P-e}
	P \ll \varepsilon \approx \rho c^2
,
\end{gather}
where $\rho \equiv m_i n_i$ is the total mass density.
The components of $\Delta T^{\mu\nu}_{({\rm EM + vortex})}$ are also much smaller than $\rho c^2$.

Below, all the three-vectors (shown in boldface) are defined in the laboratory frame.
Note that all scalar thermodynamic quantities (e.g., particle number density $n_i$) in this paper are measured in the comoving frame;
however, in the laboratory frame they have the same values in the low-velocity limit.

\vspace{0.2 cm}
\noindent
%
{\bf Nonrelativistic three-velocities}

For convenience, let us
first introduce some nonrelativistic quantities.
The four-velocity $u^\mu$ is expressed through the normal (nonsuperfluid) velocity ${\pmb V}_{\rm norm}$ of nonrelativistic hydrodynamics
by the formula
\begin{gather}
\label{eq:nonrel:u}
	u^\mu
	\equiv (u^0, {\pmb u})
	= \left( 
		\frac{1}{\sqrt{1 - \frac{{\pmb V}_{\rm norm}^2}{c^2}}},
		\frac{{\pmb V}_{\rm norm}}{c \sqrt{1 - \frac{{\pmb V}_{\rm norm}^2}{c^2}}},
	\right)
	\approx
	\left( 1, \frac{{\pmb V}_{\rm norm}}{c} \right)
.
\end{gather}
In what follows, we retain only leading-order terms in ${\pmb V}_{\rm norm} / c$ and ${\pmb V}_{{\rm s}i} / c$ in all equations.

The four-vector $w_{(i)}^\mu$ is related to the nonrelativistic superfluid velocity ${\pmb V}_{{\rm s}i}$ by \cite{ga06,gusakov16}
\begin{gather}
\label{eq:nonrel:w}
	w_{(i)}^\mu
	= m_i c V^\mu_{({\rm s}i)} - \mu_i u^\mu
,
\end{gather}

where
$ V^\mu_{({\rm s}i)}
\equiv \left( V^0_{({\rm s}i)}, {\pmb V}_{{\rm s}i} \right) $
and  $V^0_{({\rm s}i)}$
can be found from Eqs.~\eqref{eq:uw=0} and \eqref{eq:nonrel:w}:
\begin{gather}
\label{eq:nonrel:Vs0}
V^0_{({\rm s}i)} 
	= \frac{\mu_i}{m_i c u^0} + \frac{{\pmb u} {\pmb V}_{{\rm s}i}}{u^0}
.
\end{gather}
In the low-velocity limit
\begin{gather}
\label{eq:nonrel:w2}
	w_{(i)}^\mu
	= \left( w_{(i)}^0, {\pmb w}_{(i)} \right)
	\approx
	 \left(0, m_i c {\pmb V}_{{\rm s}i} - \mu_i \frac{{\pmb V}_{\rm norm}}{c} \right)
.
\end{gather}
For nonrelativistic particles 
$\mu_i \approx m_i c^2$,
and, in the fully nonrelativistic limit,
the vector ${\pmb w}_{(i)}$  reduces to
\begin{gather}
\label{eq:nonrel:w3}
	{\pmb w}_{(i)}
	= m_i c \left( {\pmb V}_{{\rm s}i} - {\pmb V}_{\rm norm} \right)
.
\end{gather}

Being expressed in terms of $V^\mu_{({\rm s}i)}$,
the vorticity tensor $\mathcal{V}_{(i)}^{\mu\nu}$ \eqref{eq:Vmunu}
reads
(recall that, starting from Sec. \ref{sec:f-dj}, we ignore viscosity and set $\varkappa_i = 0$)
\begin{gather}
	\mathcal{V}_{(i)}^{\mu\nu}
	= m_i \left[ \partial^\mu V^\nu_{({\rm s}i)} - \partial^\nu V^\mu_{({\rm s}i)} \right]
	+ \frac{e_i}{c} F^{\mu\nu}
.
\end{gather}

In the fully nonrelativistic limit it is also convenient
to introduce the nonsuperfluid particle velocities ${\pmb V}_{i}$,
in order to express the spatial part of the particle current
$j_{(i)}^\mu \equiv \left( j_{(i)}^0, {\pmb j}_i \right)$,
as a sum of nonsuperfluid and superfluid currents
(with velocities ${\pmb V}_i$ and ${\pmb V}_{{\rm s}i}$, respectively):
\begin{gather}
\label{eq:nonrel:j-Vi}
	{\pmb j}_i
	= \left( n_i - \frac{1}{m_i} \sum_{k} \rho_{ik} \right) \frac{{\pmb V}_i}{c}
	+ \frac{1}{m_i} \sum_{k} \frac{\rho_{ik} {\pmb V}_{{\rm s}k}}{c}
.
\end{gather}
Note that no summation over index $i$ is assumed
in Eqs.\ \eqref{eq:nonrel:j-Vi}--\eqref{eq:nonrel:dj-Vi-non-sfl},
and only linear terms in velocities are taken into account.
Comparing Eq.~\eqref{eq:nonrel:j-Vi} with definitions \eqref{eq:jmu}, \eqref{eq:nonrel:u}, and \eqref{eq:nonrel:w2},
one can express $\Delta {\pmb j}_i$ through ${\pmb V}_i$ as
\begin{gather}
\label{eq:nonrel:dj-Vi}
	\Delta {\pmb j}_i
	= \left( n_i - \frac{1}{m_i} \sum_{k} \rho_{ik} \right) 
		\frac{{\pmb V}_i - {\pmb V}_{\rm norm}}{c}
.
\end{gather}
For nonsuperfluid particles Eq. \eqref{eq:nonrel:dj-Vi}
reduces to 
\begin{gather}
\label{eq:nonrel:dj-Vi-non-sfl}
	\Delta {\pmb j}_i = n_i \frac{({\pmb V}_i - {\pmb V}_{\rm norm})}{c}
.
\end{gather}

Using the above definitions, below we present
the low-velocity version
of equations of Sec.~\ref{sec:mhd-full}, and also discuss how they
will be modified in
the fully nonrelativistic limit.
The full set of equations contains dynamic equations for seven unknown functions
${\pmb B}$, ${\pmb V}_{\rm norm}$, ${\pmb w}_{n}$, $n_n$, $n_e$, $n_\mu$, and $S$,
supplemented by algebraic relations allowing one to find all other quantities.

\vspace{0.2 cm}
\noindent
%
{\bf Dynamic equations}

\begin{enumerate}
\item
In the low-velocity limit
the continuity equations for neutrons \eqref{eq:full-mhd:jn}, electrons \eqref{eq:full-mhd:je} and muons \eqref{eq:full-mhd:jmu}
read
\begin{gather}
\label{eq:nonrel:cont-n}
	\frac{\partial n_n}{\partial t}
	+ {\pmb \nabla} \left[
		n_n {\pmb V}_{\rm norm} 
	+ c Y_{nk} {\pmb w}_{k}
		+ c\Delta {\pmb j}_{n} 
	\right]
	= 0
,\\
	\frac{\partial n_e}{\partial t}
	+ {\pmb \nabla} \left[
	n_e {\pmb V}_{\rm norm} 
	+ c\Delta {\pmb j}_{e} 
	\right]
	= 0
,\\
	\frac{\partial n_\mu}{\partial t}
	+ {\pmb \nabla} \left[
	n_\mu {\pmb V}_{\rm norm} 
	+ c\Delta {\pmb j}_{\mu} 
	\right]
	= 0
.
\end{gather}
In the fully nonrelativistic limit these equations can be presented,
in terms of the velocities ${\pmb V}_i$ and ${\pmb V}_{{\rm s}i}$ \cite{ab76}, as
\begin{gather}
	\frac{\partial \rho_n}{\partial t}
	+ {\pmb \nabla} \left[
		\left( \rho_n - \rho_{nn} - \rho_{np} \right){\pmb V}_n 
		+ \rho_{nk} {\pmb V}_{{\rm s}k}
	\right]
	= 0
,\\
	\frac{\partial \rho_e}{\partial t}
	+ {\pmb \nabla} \left(
	\rho_e {\pmb V}_{e} 
	\right)
	= 0
,\\
	\frac{\partial \rho_\mu}{\partial t}
	+ {\pmb \nabla} \left(
	\rho_\mu {\pmb V}_{\mu} 
	\right)
	= 0
,
\end{gather}
where $\rho_i \equiv m_i n_i$ and no summation over $i$ is assumed.

\item 
The entropy generation equation \eqref{eq:full-mhd:dS},
which is convenient to use instead of the energy conservation law, reduces to
\begin{gather}
\label{eq:nonrel:dS}
\frac{1}{c} \pd{S}{t}
+ {\pmb \nabla} \left( 
S \frac{{\pmb V}_{\rm norm}}{c}
- \frac{\mu_i}{T} \Delta {\pmb j}_{i}
\right)
= 	\frac{\mu_p n_p^2}{c^3 T}
{\pmb f}_{p} {\pmb W}_{p}
- \Delta {\pmb j}_{i} {\pmb d}_{(i)}
,
\end{gather}
and the total momentum conservation equation reads
\begin{gather}
\label{eq:nonrel:momentum-law}
	\frac{1}{c} \frac{\partial T^{0l}}{\partial t}
	+ \nabla_{m} T^{lm}
	= 0
,
\end{gather}
where the spatial indices $l$ and $m$ run over $l,m=1,2,3$,
and the energy-momentum tensor $T^{\mu\nu}$ is specified by
Eq.\ \eqref{eq:full-mhd:Tmunu}. 
In the fully nonrelativistic limit the 
momentum density $T^{0l} / c$
reduces simply to 
$T^{0l} / c = \rho V^l_{\rm norm}
	+ \sum_{ik} \rho_{ik} \left( V^l_{{\rm s}k} - V^l_{\rm norm}\right)$,
while $T^{lm}$ is given by
Eq.~\eqref{eq:nonrel:Tlm} below.
Then Eq.~\eqref{eq:nonrel:momentum-law}, with the help of the Gibbs-Duhem relation \eqref{eq:nonrel:dP},
can be represented as
\begin{multline}
	\pd{}{t} \left[
		\rho V^l_{\rm norm}
		+ \sum_{ik} \rho_{ik} \left( V^l_{{\rm s}k} - V^l_{\rm norm}\right)
	\right]
	+ \nabla^m
		\left[
			\rho V_{\rm norm}^l V_{\rm norm}^m
			+ \sum_{ik} \rho_{ik} \left( V_{{\rm s}i}^l V_{{\rm s}k}^m - V_{\rm norm}^l V_{\rm norm}^m \right)
		\right]
	=\\ - n_i \nabla^l \mu_i - S \nabla^l T
		+ \rho_{ik} \nabla^l
		    \left[
			\frac{\left( {\pmb V}_{{\rm s}i} - {\pmb V}_{\rm norm} \right)
				  \left( {\pmb V}_{{\rm s}k} - {\pmb V}_{\rm norm} \right)}
				 {2}
		    \right]
	  - \frac{1}{4\pi} \left[ {\pmb B} \times {\rm curl} \left( H_{\rm c1} {\pmb b}\right) \right]^l
.
\end{multline}
Here the last term in the right-hand side describes buoyancy and tension forces acting on proton flux tubes.
This term replaces the Lorentz term
${\pmb J}_{\rm free} \times {\pmb B}$
of the ordinary MHD, which vanishes due to the screening of electric current inside the superconductor
(see, e.g., Refs. \cite{ep77,gas11}\footnote{%
Note that the force $F_{\rm mag}^i$ in Eq.~(95) of Ref.~\cite{gas11}
contains an additional term, $-(\rho_p)/(4\pi) \nabla^i \left( B \partial H_c / \rho \right)$;
in our formulation this term is included in ${\pmb \nabla} \mu_i$ due to renormalization of the chemical potential,
see equation~(G25) in GD16.
}).

\item
Superfluid equation \eqref{eq:full-mhd:sfl-n}, written for neutrons in the absence of vortices,
in the three-dimensional form reduces to the two equations,
\begin{gather}
\label{eq:nonrel:sfl-n-0}
	\frac{1}{c} \pd{{\pmb V}_{{\rm s}n}}{t} + {\pmb \nabla} V_{{\rm s}n}^0 = 0
,\\
	{\rm curl} {\pmb V}_{{\rm s}n} = 0
,
\end{gather}
where $V_{{\rm s}n}^0$ is given by Eq.~\eqref{eq:nonrel:Vs0}.
One can also obtain a nonrelativistic version of Eq.~\eqref{eq:nonrel:sfl-n-0},
assuming that velocities are small and neutrons are nonrelativistic (see Ref.~\cite{gusakov16}, Appendix C):
\begin{gather}
\label{eq:nonrel:sfl-n}
	\pd{{\pmb V}_{{\rm s}n}}{t} + ({\pmb V}_{{\rm s}n}{\pmb \nabla}){\pmb V}_{{\rm s}n} 
	+ {\pmb \nabla}\left[ \breve{\mu}_n 
	- \frac{1}{2} \left| {\pmb V}_{{\rm s}n}-{\pmb V}_{\rm norm}\right|^2\right]
= 0
,
\end{gather}
where $\breve{\mu}_n \equiv (\mu_n - m_n c^2)/m_n$.

\item
The ``magnetic evolution'' equation [the same as equation (I23) in GD16] is obtained from Maxwell equation
\begin{gather}
	{\rm curl} {\pmb E}=- \frac{1}{c} \pd{\pmb B}{t}
\end{gather}
by substituting ${\pmb E}$ from Eq.~\eqref{eq:nonrel:sfl-p-0}
(see below)
and neglecting the terms depending on 
${\rm curl} \,{\pmb V}_{{\rm s}p}$ in comparison 
to the similar terms depending on $e_p/(m_p c) \, {\pmb B}$:
\begin{equation}
\label{eq:nonrel:dBdt}
\frac{\partial {\pmb B}}{\partial t} + {\rm curl} 
\left(
\frac{\mu_p n_p}{e_p c} \, {\pmb f}_p + {\pmb B}\times{\pmb V}_{\rm norm}
\right)=0.
\end{equation}
\end{enumerate}

The above equations describe time evolution of magnetic field ${\pmb B}$,
velocities ${\pmb V}_{\rm norm}$ and ${\pmb V}_{{\rm s}n}$ (or, equivalently, ${\pmb w}_n$),
as well as scalar thermodynamic quantities ($n_i$ and $S$).
Note that the superfluid velocity for protons,
${\pmb V}_{{\rm s}p}$ (or ${\pmb w}_p$),
is expressed from the screening condition \eqref{eq:nonrel:screening},
and thus does not provide an additional dynamic degree of freedom;
the diffusive currents $\Delta {\pmb j}_{i}$ (or velocities ${\pmb V}_i$ of nonsuperfluid components) are also expressed algebraically via Eq.~\eqref{eq:nonrel:dj}.

\vspace{0.2 cm}
\noindent
%
{\bf Algebraic relations}

\begin{enumerate}
\item
In the low-velocity limit 
the small quantity $w_{(i)}^\mu w_{(k)\mu}$ that enters
the thermodynamic relations \eqref{eq:mhd-limit:energy}--\eqref{eq:mhd-limit:dP}
reduces to ${\pmb w}_{(i)} {\pmb w}_{(k)}$ [see Eq.~\eqref{eq:nonrel:w2}].
As a result, any thermodynamic quantity (e.g., the energy density $\varepsilon$) should be expressed as functions of the variables
$n_i$, $S$, ${\pmb w}_{(i)} {\pmb w}_{(k)}$, and $B$,
\begin{gather}
\label{eq:nonrel:energy}
	\varepsilon = \varepsilon \left( n_i, S, {\pmb w}_{(i)} {\pmb w}_{(k)}, B \right)
,
\end{gather}
whereas the second law of thermodynamics and the Gibbs-Duhem relation read, respectively,
\begin{gather}
\label{eq:nonrel:2ndlaw}
	d \varepsilon = \mu_i \, dn_i 
		+ T \, dS 
		+ \frac{Y_{ik}}{2} \, d \left( {\pmb w}_{(i)} {\pmb w}_{(k)}  \right)
		+ \frac{1}{4\pi} H_{\rm c1} dB
,\\
\label{eq:nonrel:dP}
	dP = n_i  \, d\mu_i + S \, dT
	- \frac{Y_{ik}}{2} \, d \left( {\pmb w}_{(i)} {\pmb w}_{(k)}  \right)
	- \frac{1}{4\pi} H_{\rm c1} dB
.
\end{gather}
In the fully nonrelativistic limit the term
$\frac{Y_{ik}}{2} \, d \left( {\pmb w}_{(i)} {\pmb w}_{(k)}  \right)$
reduces, in view of Eqs.~\eqref{eq:nonrel:rho-ik} and \eqref{eq:nonrel:w3}, to
\begin{gather}
	\frac{Y_{ik}}{2} \, d \left( {\pmb w}_{(i)} {\pmb w}_{(k)}  \right)
	= \rho_{ik}
		d \frac{\left( {\pmb V}_{{\rm s}i} - {\pmb V}_{\rm norm} \right)
		       \left( {\pmb V}_{{\rm s}k} - {\pmb V}_{\rm norm} \right)
		       }{2}		       
.
\end{gather}

\item
Proton number density $n_p$ and superfluid proton velocity ${\pmb V}_{{\rm s}p}$ can be found
from the quasineutrality \eqref{eq:full-mhd:quasineutrality}
and screening \eqref{eq:full-mhd:screening} conditions,
\begin{gather}
\label{eq:nonrel:quasineutrality}
	n_p = n_e + n_\mu
	,\\
\label{eq:nonrel:screening}
	{\pmb j}_p - {\pmb j}_e - {\pmb j}_\mu
	= 
	Y_{pk} {\pmb w}_{k}
	+ \left( \Delta {\pmb j}_{p}  - \Delta {\pmb j}_{e} - \Delta {\pmb j}_{\mu} \right) 
	= 0
.
\end{gather}
For nonrelativistic matter the screening condition \eqref{eq:nonrel:screening},
written in terms of ${\pmb V}_i$ and ${\pmb V}_{{\rm s}p}$, takes the form
\begin{gather}
\label{eq:nonrel:screening-2}
	\frac{\rho_{pk}}{m_p}
		\left( {\pmb V}_{{\rm s}k} - {\pmb V}_{p} \right)
	+ n_p {\pmb V}_{p} - n_e {\pmb V}_{e} - n_\mu {\pmb V}_{\mu} 
	= 0
.
\end{gather}

\item
The energy-momentum tensor $T^{\mu\nu}$, employed in Eq.~\eqref{eq:nonrel:momentum-law},
is specified by Eqs.~\eqref{eq:full-mhd:Tmunu} and \eqref{eq:full-mhd:TVM-2}.
In the fully nonrelativistic limit its
spatial part
$T^{lm}$ ($l,m = 1,2,3$), with the help of relations 
\eqref{eq:nonrel:rho-ik}--\eqref{eq:nonrel:u}, and \eqref{eq:nonrel:w3},
reduces to
[cf. Ref.~\cite{ab76} and equation~(I22) in GD16]
\begin{gather}
\label{eq:nonrel:Tlm}
	T^{lm}
	= \left( \rho - \sum_{ik} \rho_{ik} \right) V_{\rm norm}^l V_{\rm norm}^m
	+ \sum_{ik} \rho_{ik} V_{{\rm s}i}^l V_{{\rm s}k}^m
	+ P \delta^{lm}
	+ \frac{H_{c1}}{4\pi} \left( B \delta^{lm} - \frac{B^l B^m}{B} \right)
.	
\end{gather}

\item 
$\Delta {\pmb j}_{i}$ and ${\pmb f}_{p}$
are expressed through ${\pmb d}_{k}$ and ${\pmb W_p}$ [see Eqs.~\eqref{eq:full-mhd:fp} and \eqref{eq:full-mhd:dj}]:
\begin{gather}
\label{eq:nonrel:fp}
\begin{split}
	- \frac{\mu_p n_p^2}{c^3 T} {\pmb f}_{p}
= &- \mathcal{A}_{pp}^\parallel {\pmb W}_{p\parallel}
- \mathcal{A}_{pp}^\perp {\pmb W}_{p\perp}
- \mathcal{A}_{pp}^H \left[ {\pmb W}_{p\perp} \times {\pmb b} \right]
\\&
- \mathcal{B}_{pk}^{\parallel} {\pmb d}_{k\parallel}
- \mathcal{B}_{pk}^{\perp} {\pmb d}_{k\perp}
- \mathcal{B}_{pk}^{H} \left[ {\pmb d}_{k\perp} \times {\pmb b} \right]
.
\end{split}
\\
\label{eq:nonrel:dj}
\begin{split}
\Delta {\pmb j}_{i}
= &-\mathcal{C}_{ip}^{\parallel} {\pmb W}_{p\parallel}
- \mathcal{C}_{ip}^{\perp}  {\pmb W}_{p\perp}
- \mathcal{C}_{ip}^{H} \left[ {\pmb W}_{p\perp} \times {\pmb b} \right]
\\&
- \mathcal{D}_{ik}^{\parallel} {\pmb d}_{k\parallel}
- \mathcal{D}_{ik}^{\perp} {\pmb d}_{k\perp}
- \mathcal{D}_{ik}^{H} \left[ {\pmb d}_{k\perp} \times {\pmb b} \right]
,
\end{split}
\end{gather}
where
\begin{gather}
	{\pmb d}_{k\parallel} \equiv \left({\pmb d}_{k}  {\pmb b} \right) {\pmb b}
,\quad
	{\pmb d}_{k\perp} \equiv {\pmb d}_k -  \left({\pmb d}_{k}  {\pmb b} \right) {\pmb b}
,\quad
	{\pmb W}_{p\parallel} \equiv \left({\pmb W}_{p}  {\pmb b} \right) {\pmb b}
,\quad
	{\pmb W}_{p\perp} \equiv {\pmb W}_p -  \left({\pmb W}_{p}  {\pmb b} \right) {\pmb b}
,\\
	{\pmb b} \equiv \frac{{\pmb B}}{B}
,\\
\label{eq:nonrel:d}
	{\pmb d}_{k} =
	{\pmb \nabla} \left(\frac{\mu_k}{T}\right) 
	- \frac{e_k}{T}  
		\left[{\pmb E} + \frac{{\pmb V}_{\rm norm}}{c} \times {\pmb B} \right] 
,\\
\label{eq:nonrel:Wp}
	{\pmb W}_p 
	=
		\frac{c Y_{pk}}{n_p} {\pmb w}_{(k)}
		+ \frac{c}{4\pi e_p n_p} 
			{\rm curl} \left( H_{\rm c1}\, {\pmb b}\right)
.
\end{gather}

\item
The electric field ${\pmb E}$ is expressed
from  the superfluid equation \eqref{eq:full-mhd:sfl-p} for protons, 
\begin{gather}
\label{eq:nonrel:sfl-p-0}
\pd{{\pmb V}_{{\rm s}p}}{t}
	+ c {\pmb \nabla} V_{{\rm s}p}^0 
	+ {\rm curl}{\pmb V}_{{\rm s}p}\times {\pmb V}_{\rm norm} 
	=
	- \frac{\mu_p n_p}{m_p c^2} \, {\pmb f}_p
	+ \frac{e_p}{m_p} \, \left({\pmb E+\frac{{\pmb V}_{\rm norm}}{c}\times {\pmb B}}\right)
,
\end{gather}
which, in the nonrelativistic limit,
takes the form
[cf. GD16, Eq.\ (I7)]
\begin{eqnarray}
\label{eq:nonrel:sfl-p}
\pd{{\pmb V}_{{\rm s}p}}{t}
	+ ({\pmb V}_{{\rm s}p}{\pmb \nabla}){\pmb V}_{{\rm s}p} 
	+ {\pmb \nabla}\left[ \breve{\mu}_p 
	- \frac{1}{2} \left| {\pmb V}_{{\rm s}p}-{\pmb V}_{\rm norm}\right|^2
	\right]
	&=&-{\rm curl}{\pmb V}_{{\rm s}p}\times \left({\pmb V}_{\rm norm}-{\pmb V}_{{\rm s}p}\right) 
	\nonumber\\
	&-& n_p \, {\pmb f}_p
	+ \frac{e_p}{m_p} \, \left({\pmb E+\frac{{\pmb V}_{\rm norm}}{c}\times {\pmb B}}\right)
,
\end{eqnarray}
where $\breve{\mu}_p \equiv (\mu_p-m_p c^2)/m_p$.

Note that the right-hand sides of Eqs.~\eqref{eq:nonrel:fp} and \eqref{eq:nonrel:dj}
implicitly contain $\Delta {\pmb j}_{i}$
and ${\pmb E}$ (see footnote~\ref{footnote:implicitly-contains}),
therefore one has to solve equations
\eqref{eq:nonrel:fp}, \eqref{eq:nonrel:dj}, and \eqref{eq:nonrel:sfl-p}
{\it simultaneously} in order to obtain
closed-form
expressions for ${\pmb f}_p$, $\Delta {\pmb j}_{i}$, and ${\pmb E}$.
\end{enumerate}
%

\vspace{0.2 cm}
\noindent
%
{\bf Remark 1.} 
If neutrons
and
protons are completely superfluid, then $\Delta {\pmb j}_{n}$
and
$\Delta {\pmb j}_{p}$
(which describe dissipative corrections to the nonsuperfluid currents)
vanish together with the
corresponding transport coefficients.

\vspace{0.2 cm}
\noindent
%
{\bf Remark 2.}
The magnetic evolution equation \eqref{eq:nonrel:dBdt} 
can be further simplified if
transport coefficients $\mathcal{A}_{pp}^\parallel$ and $\mathcal{B}_{pi}^\parallel$
in Eq.~\eqref{eq:nonrel:fp} are small.
Then ${\pmb f}_p$ can be presented as
\begin{gather}
\label{eq:nonrel:fp-VL}
	{\pmb f}_p
	= \frac{e_p c}{\mu_p n_p} 
		\left[ {\pmb B} \times \left( {\pmb V}_{{\rm L}p} - {\pmb V}_{\rm norm} \right) \right]
,
\end{gather}
where
\begin{gather}
\label{eq:nonrel:vLp}
{\pmb V}_{{\rm L}p}
=	{\pmb V}_{\rm norm} - \frac{c^2 T}{e_p n_p B}
	\left( \mathcal{A}_{pp}^H {\pmb W}_{p} + \mathcal{B}_{pk}^H {\pmb d}_{k} \right)
	+ \frac{c^2 T}{e_p n_p B}
	\left( \mathcal{A}_{pp}^\perp {\pmb W}_{p} + \mathcal{B}_{pk}^\perp {\pmb d}_{k} \right) \times {\pmb b}
\end{gather}
is the nonrelativistic velocity of proton vortices
[spatial part of the four-vector $v^{\mu}_{({\rm L}p)}$ multiplied by $c$, see Eq.~\eqref{eq:full-mhd:vLp}].
Eq.~\eqref{eq:nonrel:dBdt} can then be rewritten in the form
[cf. GD16, equation (I24)]
\begin{equation}
    \frac{\partial {\pmb B}}{\partial t} + {\rm curl} \,
    ({\pmb B}\times {\pmb V}_{{\rm L}p})=0,
\end{equation}
which simply states that the magnetic field is transferred by the vortices.

\vspace{0.2 cm}
\noindent
%
{\bf{Remark 3.}}

One can easily account for the presence of neutron vortices,
provided that we neglected
their effect on diffusion and ignore vortex-flux tube interaction
(see Sec.~\ref{sec:special-cases:sfl-nV}).
Under these assumptions,
all equations of this section remain the same,
except for Eqs.~\eqref{eq:nonrel:sfl-n-0}--\eqref{eq:nonrel:sfl-n}, 
which should be replaced with 
\begin{gather}
\label{eq:nonrel:sfl-n-vortices}
\pd{{\pmb V}_{{\rm s}n}}{t}
	+ c {\pmb \nabla} V_{{\rm s}n}^0 
	+ {\rm curl}{\pmb V}_{{\rm s}n}\times {\pmb V}_{\rm norm} 
	=
	- \frac{\mu_n n_n}{m_n c^2} \, {\pmb f}_n
,
\end{gather}
and Eq.~\eqref{eq:nonrel:dS},
which should be replaced with
\begin{gather}
    \frac{1}{c} \pd{S}{t}
    + {\pmb \nabla} \left( 
    S \frac{{\pmb V}_{\rm norm}}{c}
    - \frac{\mu_i}{T} \Delta {\pmb j}_{i}
    \right)
    = 	\frac{\mu_p n_p^2}{c^3 T}
    {\pmb f}_{p} {\pmb W}_{p}
    + \frac{\mu_n n_n^2}{c^3 T}
    {\pmb f}_{n} {\pmb W}_{n}
    - \Delta {\pmb j}_{i} {\pmb d}_{(i)}
,
\end{gather}
where ${\pmb W}_{n}$ is given by
[see Eq.~\eqref{eq:mhd-limit:Wn}]
\begin{gather}
    {\pmb W}_{n}
	\equiv \frac{1}{n_n}
	        c Y_{nk} {\pmb w}_{(k)} 
,
\end{gather}
and
${\pmb f}_{n}$
is [see Eq.~\eqref{eq:sfl-nV:fn}]
\begin{gather}
\label{eq:nonrel:fn}
	- \frac{\mu_n n_n^2}{c^3 T} {\pmb f}_{n}
	= - \mathcal{A}_{nn}^\parallel {\pmb W}_{n\parallel}
	  - \mathcal{A}_{nn}^\perp {\pmb W}_{n\perp}
	  - \mathcal{A}_{nn}^H \left[ {\pmb W}_{n\perp} \times {\pmb \omega} \right]
,\\
	{\pmb W}_{n\parallel} \equiv \left({\pmb W}_{n}  {\pmb \omega} \right) {\pmb \omega}
,\quad
	{\pmb W}_{n\perp} \equiv {\pmb W}_n -  \left({\pmb W}_{n}  {\pmb \omega} \right) {\pmb \omega}
,\quad
	{\pmb \omega} \equiv
		\frac{{\mathbfcal{V}}_{({\rm M}n)}}
			 {{\mathcal{V}}_{({\rm M}n)}}
.
\end{gather}
In the nonrelativistic limit Eq.~\eqref{eq:nonrel:sfl-n-vortices} reduces to [cf. GD16, Eq.\ (I7)]
\begin{gather}
	\pd{{\pmb V}_{{\rm s}n}}{t}
	+ ({\pmb V}_{{\rm s}n}{\pmb \nabla}){\pmb V}_{{\rm s}n} 
	+ {\pmb \nabla}\left[ \breve{\mu}_n 
		- \frac{1}{2} \left| {\pmb V}_{{\rm s}n}-{\pmb V}_{\rm norm}\right|^2
	\right]
	= -{\rm curl}{\pmb V}_{{\rm s}n}\times \left({\pmb V}_{\rm norm}-{\pmb V}_{{\rm s}n}\right) 
	  - n_n \, {\pmb f}_n
.
\end{gather}

\section{
Phenomenological transport coefficients
in the low-temperature limit}
\label{sec:mf}

Here we establish a connection between our transport coefficients
and the
mutual friction
parameters/momentum transfer rates
of microscopic theory.
To this aim, we analyze the
equation of motion for individual proton vortices,
as well as the Euler-like equations for nonsuperfluid particles in the $npe\mu$ matter.%
%
\footnote{These Euler-like equations follow from the transport equations written for each particle species,
see, e.g., Refs.\ \cite{gr92,ys91a,dgs20, kgk21}.
}
We present an algorithm that allows us to
find microscopic expressions for $\Delta {\pmb j}_i$ and ${\pmb V}_{{\rm L}p}$,
compare them with the phenomenological equations \eqref{eq:nonrel:dj} and \eqref{eq:nonrel:vLp},
and, finally, obtain the expressions for the phenomenological transport coefficients
$\mathcal{A}_{ik}^{\mu\nu}$, $\mathcal{B}_{ik}^{\mu\nu}$, $\mathcal{C}_{ik}^{\mu\nu}$, and $\mathcal{D}_{ik}^{\mu\nu}$.

As in Appendix~\ref{sec:mhd-full-nonrel}, we work in the MHD limit, ignore neutron vortices, and assume that all macroscopic velocities are nonrelativistic.
For the sake of simplicity, we further make some additional assumptions.
Namely, we adopt the low-temperature limit ($T \to 0$),
ignore all the terms depending on ${\pmb \nabla} T$,
and assume that protons and neutrons are completely superfluid (no Bogoliubov thermal excitations),
so that only electrons and muons can scatter off the vortex cores.
In addition, we also neglect entrainment
between superfluid neutrons and protons, i.e., set $Y_{np} = 0$.

The proton vortex velocity ${\pmb V}_{{\rm L}p}$
enters the
equation describing the balance of forces acting on a proton vortex.
Neglecting small vortex mass, the latter equation
takes the form \cite{gusakov19}
\begin{gather}
\label{eq:mf:F=0}
	\sum_{i=e,\mu}
	{\pmb F}_{i \to V} + {\pmb F}_{\rm ext} = 0
,
\end{gather}
where
\begin{gather}
\label{eq:mf:Fj}
	{\pmb F}_{i \to V}
	= - D_i \left[ 
		{\pmb b} \times \left[ 
			{\pmb b} \times \left( {\pmb V}_i - {\pmb V}_{{\rm L}p} \right) 
			\right]
		\right]
		+ D'_i \left[ 
			{\pmb b} \times \left( {\pmb V}_i - {\pmb V}_{{\rm L}p} \right) 
		\right]
\end{gather}
is the velocity-dependent force per unit length acting on a vortex from particle species $i$,
${\pmb V}_i \equiv c {\pmb j}_i / n_i$
is the velocity of particle species $i$,
and coefficients $D_i$ and $D'_i$ are calculated from microphysics (see Ref.~\cite{gusakov19} and references therein).
In the absence of diffusion
the phenomenological mutual friction parameters $\alpha_p$, $\beta_p$, and $\gamma_p$ employed in GD16 can be expressed through  $D_i$ and $D'_i$ as
\begin{gather}
    \frac{\mu_p n_p}{c^2} \alpha_p
    = \frac{\pi \hbar n_p  (D'_e + D'_\mu)}
           {(D_e + D_\mu)^2 + (D'_e + D'_\mu)^2}
,\\
    \frac{\mu_p n_p}{c^2} \beta_p
    = \frac{\pi \hbar n_p  (D_e + D_\mu)}
           {(D_e + D_\mu)^2 + (D'_e + D'_\mu)^2}
,\\
    \gamma_p = 0
.
\end{gather}
To obtain these relations, one has to solve Eq.~\eqref{eq:mf:F=0} with ${\pmb V}_e = {\pmb V}_\mu = {\pmb V}_{\rm norm}$ and compare the result with equations (101) and (I25) of GD16.

The first and the second term in Eq.~\eqref{eq:mf:Fj} describe the (dissipative) drag force and the (nondissipative) transverse force, respectively.
${\pmb F}_{\rm ext}$ is the velocity-independent force per unit length;
it is the
sum of buoyancy and tension forces \cite{dg17}:
\begin{gather}
\label{eq:mf:Fext}
	{\pmb F}_{\rm ext}
	= - \frac{\hbar c}{4 e_p} \left[ {\pmb b} \times {\rm curl} \left( H_{\rm c1} {\pmb b} \right) \right]
.
\end{gather}
Using Eqs. \eqref{eq:nonrel:screening} and \eqref{eq:nonrel:Wp},
and noting that $\Delta {\pmb j}_{p} = 0$ (since all protons are superconducting),
one can present ${\pmb F}_{\rm ext}$ as
\begin{gather}
\label{eq:mf:Fext-2}
	{\pmb F}_{\rm ext}
	= - \pi \hbar n_p \left[ {\pmb b} \times
					 \left(
					 	{\pmb W}_p
					 	- \frac{c}{n_p} \Delta {\pmb j}_e
					 	- \frac{c}{n_p} \Delta {\pmb j}_\mu
					\right)
	\right]
.
\end{gather}

The velocities ${\pmb V}_e$ and ${\pmb V}_\mu$ can be found from
the Euler equations \cite{gko20} ($i=e,\mu$ and no summation over $i$ is assumed)
\begin{gather}
\label{eq:mf:euler}
	n_i	\left[\pd{}{t} + \left({\pmb V}_i {\pmb \nabla}  \right)  \right]
		\left( \frac{\mu_i}{c^2} {\pmb V}_i \right)
	= 
	- n_i {\pmb \nabla} \mu_i
	- \frac{\mu_i n_i}{c^2} {\pmb \nabla} \phi
	- \sum_{k \neq i} J_{ik} \left( {\pmb V}_i - {\pmb V}_k \right)
	- N_{{\rm V}p} {\pmb F}_{i \to V}
,
\end{gather}
where $\phi$ is the gravitational potential,
$J_{ik} = J_{ki}$ is the momentum transfer rate per unit volume between particle species $i$ and $k$,
and
$N_{{\rm V}p} = B / \hat{\phi}_{p0} = e_p B / (\pi \hbar c)$
is the number of proton vortices per unit area.
The Lorentz force is contained in the last term in the right-hand side of Eq.~\eqref{eq:mf:euler},
since we assume that all the electromagnetic field is generated by proton vortices.
Note that, e.g., in the similar equations of Ref.~\cite{gko20}
the vector ${\pmb {\mathcal{F}}_{pi}}$ from this reference
includes only the drag force [the second term in Eq.~\eqref{eq:mf:Fj}],
whereas the Lorentz force [the first term in Eq.~\eqref{eq:mf:Fj}] is written out separately.

Since in the hydrodynamic regime the velocities ${\pmb V}_i$
are close to one another, one can simplify the left-hand side of Eq.~\eqref{eq:mf:euler}
by replacing ${\pmb V}_i$ 
with the average mass velocity of nonsuperfluid particles
${\pmb U} \equiv \left( \mu_e n_e {\pmb V}_e + \mu_\mu n_\mu {\pmb V}_\mu \right) / 
( \mu_e n_e+$ $ \mu_\mu n_\mu 
)$ \cite{brag65,ys91a},
which, in the low-temperature limit,
coincides with ${\pmb V}_{\rm norm}$ introduced in Eq.~\eqref{eq:nonrel:u}.
Below we work in the comoving frame, specified by the condition ${\pmb V}_{\rm norm} = 0$,
or, in terms of ${\pmb V}_e$ and ${\pmb V}_\mu$,
\begin{gather}
\label{eq:mf:mnu=0}
\mu_e n_e {\pmb V}_e + \mu_\mu n_\mu {\pmb V}_\mu = 0
.
\end{gather}
The left-hand side of Eq.~\eqref{eq:mf:euler} in this frame reduces to $\left(\mu_i n_i/c^2 \right) \partial {\pmb U} / \partial t$.
Then, subtracting Euler equations \eqref{eq:mf:euler} (divided by $\mu_i n_i$) for electrons and muons, we obtain:
\begin{gather}
\label{eq:mf:euler-e-mu}
- \frac{{\pmb \nabla} \mu_e}{\mu_e} + \frac{{\pmb \nabla} \mu_\mu}{\mu_\mu}
- \left( \frac{1}{\mu_e n_e} + \frac{1}{\mu_\mu n_\mu} \right)
J_{e\mu} \left( {\pmb V}_e - {\pmb V}_\mu \right)
- \frac{1}{\mu_e n_e} N_{{\rm V}p} {\pmb F}_{e \to V}
+ \frac{1}{\mu_\mu n_\mu} N_{{\rm V}p} {\pmb F}_{\mu \to V}
= 0
.
\end{gather}

The set of linear algebraic equations \eqref{eq:mf:F=0}, \eqref{eq:mf:mnu=0}, and \eqref{eq:mf:euler-e-mu}
allows one to find the quantities ${\pmb V}_{{\rm L}p}$, $\Delta {\pmb j}_e$, and $\Delta {\pmb j}_\mu$.
To express them through ${\pmb W}_p$, ${\pmb d}_e$, and ${\pmb d}_\mu$,
one has to make the following substitutions in these equations:
\begin{enumerate}
	\item
	substitute ${\pmb F}_{i \to V}$ and ${\pmb F}_{\rm ext}$ from Eqs.~\eqref{eq:mf:Fj} and \eqref{eq:mf:Fext-2};
	\item
	replace ${\pmb \nabla} \mu_i$ with $T {\pmb d}_i  + e_i {\pmb E}$ [see Eq.~\eqref{eq:nonrel:d}; recall that we ignore the terms depending on ${\pmb \nabla} T$];
	\item
	replace	${\pmb E}$ with $\left(- 1/c \right) {\pmb V}_{{\rm L}p} \times {\pmb B}$
	[this condition follows from the assumption that the electric field is generated only by the vortex motion, see equation~(G15) in GD16];
	\item
	replace	${\pmb V}_i$ with $c \Delta {\pmb j}_i / n_i$ (note that we work in the comoving frame, ${\pmb V}_{\rm norm} = 0$).
\end{enumerate}
Then, solving the system of equations \eqref{eq:mf:F=0}, \eqref{eq:mf:mnu=0}, and \eqref{eq:mf:euler-e-mu},
and comparing the results with Eqs.\ \eqref{eq:nonrel:dj} and \eqref{eq:nonrel:vLp},
one can determine the coefficients
$\mathcal{A}_{pp}^{\mu\nu}$, $\mathcal{B}_{pk}^{\mu\nu}$,
$\mathcal{C}_{ip}^{\mu\nu}$, and $\mathcal{D}_{ik}^{\mu\nu}$
and directly check 
that the Onsager relations (\ref{eq:oB:Aik=Aki-2}), (\ref{eq:oB:Cik=Bki-2}), and (\ref{eq:oB:Dik=Dki-2}) are satisfied.

Since the resulting expressions are very lengthy, we do not provide them for the most general case.
Instead, we write them out in the limit
$J_{e\mu} \ll N_{{\rm V}p} D_i \ll \left| N_{{\rm V}p} D'_i \right| $,
which is realistic for typical NS conditions (see, e.g., Fig.~1 in Ref.~\cite{gko20}).
We also set $D'_i = - \pi \hbar n_i$, as argued in Refs.~\cite{gas11,gusakov19}.
Then the transport coefficients have, up to the first order in
$J_{e\mu} / \left| N_{{\rm V}p} D'_i \right| = (c J_{e\mu}) / (e_p n_i B)$
and
$D_i / \left| D'_i \right| = D_i / ( \pi \hbar n_i)$,
the following form:\footnote{
Note that the expression
\eqref{eq:mf:Bpe-perp}
for $\mathcal{B}_{pe}^\perp$ is of the second order in the small parameter $(D_i / D'_i)$; we write it down to emphasize that, generally, it does not vanish.
We also point out that we retain 
the (small) second term 
$\propto J_{e\mu}$
in the intermediate equality
in
\eqref{eq:mf:Demu-perp},
because only this term survives    
in the expression for $\mathcal{D}_{e\mu}^\perp$ 
in the nonsuperfluid MHD of DGS20.
}
\begin{gather}
\mathcal{A}_{pp}^\parallel = 0
,\\
\mathcal{A}_{pp}^\perp = \frac{e_p B \left( D_e + D_\mu \right)}{\pi \hbar c^2 T}
,\\
\mathcal{A}_{pp}^H = - \frac{e_p n_p B}{c^2 T}
,\\
\mathcal{B}_{pk}^\parallel = 0 
,\\
\label{eq:mf:Bpe-perp}
\begin{split}
\mathcal{B}_{pe}^\perp
	= \frac{
		\mu_{\mu }  \left( \mu_{\mu } n_{\mu } D_e-\mu_e n_e D_\mu \right)
			\left[n_e D_\mu \left(\mu_e^2 n_e^2+2 \mu_e^2 n_e n_{\mu }+\mu_{\mu }^2 n_{\mu }^2\right)
				+ n_{\mu } D_e  \left(\mu_e^2 n_e^2+2 \mu_{\mu }^2 n_e n_{\mu }+\mu_{\mu }^2 n_{\mu }^2\right)
			\right]
		}
		{\pi ^2 \hbar^2 c n_e n_{\mu }
			\left(n_e+n_{\mu }\right)
			\left(\mu_e^2 n_e+\mu_{\mu }^2 n_{\mu }\right)^2}
	\\+
	\frac{\mu_{\mu }  J_{e\mu} \left(\mu_e n_e+\mu_{\mu } n_{\mu }\right)^2
			\left(\mu_{\mu } n_{\mu } D_e -\mu_e n_e D_\mu\right)}
		{\pi \hbar e_p B n_e n_{\mu }
			\left(\mu_e^2 n_e+\mu_{\mu }^2 n_{\mu }\right)^2}
	\approx 0
,
\end{split}
\\
\mathcal{B}_{p\mu}^\perp = - \frac{\mu_{e}}{\mu_{\mu}} \mathcal{B}_{pe}^\perp
,\\
\mathcal{B}_{pe}^H = \frac{\mu_{\mu} \left(\mu_e n_e D_{\mu} -\mu_{\mu} n_{\mu} D_e \right)  }
							{\pi \hbar c \left( \mu_e^2 n_e + \mu_{\mu }^2 n_{\mu } \right)}
,\quad
\mathcal{B}_{p\mu}^H = - \frac{\mu_{e}}{\mu_{\mu}} \mathcal{B}_{pe}^H
,\\
\mathcal{C}_{ip}^\parallel = 0 
,\quad
\mathcal{C}_{ip}^\perp = - \mathcal{B}_{pi}^\perp
,\quad
\mathcal{C}_{ip}^H = - \mathcal{B}_{pi}^H
,\\
\mathcal{D}_{e\mu}^\parallel
= \mathcal{D}_{\mu e}^\parallel
= - \frac{\mu_e \mu_\mu n_e^2 n_\mu^2 T}
{c J_{e\mu} \left( \mu_e n_e + \mu_\mu n_\mu \right)^2}
,\\
\label{eq:mf:Demu-perp}
\mathcal{D}_{e\mu}^\perp
= \mathcal{D}_{\mu e}^\perp
=
	-\frac{\mu _e \mu_{\mu } T  \left(\mu_e^2 n_e^2 D_{\mu } + \mu _{\mu }^2 n_{\mu }^2 D_e \right)}
		{\pi \hbar e_p B \left(\mu_e^2 n_e+\mu _{\mu }^2 n_{\mu }\right)^2}
	-\frac{c \mu_e \mu_{\mu } T  J_{{e\mu }} \left(\mu_e n_e+\mu_{\mu } n_{\mu }\right)^2}
		{e_p^2 {B}^2  \left(\mu_e^2 n_e+\mu_{\mu }^2 n_{\mu }\right)^2}
\approx
	-\frac{\mu _e \mu_{\mu } T  \left(\mu_e^2 n_e^2 D_{\mu } + \mu _{\mu }^2 n_{\mu }^2 D_e \right)}
		{\pi \hbar e_p B \left(\mu_e^2 n_e+\mu _{\mu }^2 n_{\mu }\right)^2}
,\\
\mathcal{D}_{e\mu}^H
	= \mathcal{D}_{\mu e}^H
	= \frac{\mu _e \mu _{\mu } n_e n_{\mu } T}{e_p B \left(\mu _e^2 n_e +\mu _{\mu }^2 n_{\mu}\right) }
,\\
\mathcal{D}_{ee}^{\parallel,\perp,H}
= - \frac{\mu_\mu}{\mu_e} \mathcal{D}_{e\mu}^{\parallel,\perp,H}
,\quad
\mathcal{D}_{\mu\mu}^{\parallel,\perp,H}
= - \frac{\mu_e}{\mu_\mu} \mathcal{D}_{e\mu}^{\parallel,\perp,H}
.
\end{gather}

A number of comments regarding these equations is in order:
\begin{enumerate}
\item
The coefficients $\mathcal{A}_{pp}^\parallel$ and $\mathcal{B}_{pi}^\parallel$ vanish since there is no force acting along the vortex line in Eq.~\eqref{eq:mf:F=0}.
\item
$\mathcal{A}_{pp}^H$ and $\mathcal{A}_{pp}^\perp$ do not depend, in the leading order,
on the electron-muon momentum transfer rate $J_{e\mu}$;
these coefficients are proportional to, respectively,
the mutual friction
parameters
$\alpha_p$ and $\beta_p$ of nondiffusive hydrodynamics \cite{gd16} [cf. Eq.~\eqref{eq:Ann}].
Note, however, that generally \textit{all} coefficients, except for $\mathcal{D}_{ik}^\parallel$,  depend on both $J_{e\mu}$ and $D_i$.
\item
The cross-coefficient $\mathcal{B}_{pi}^H$,
which describes force acting on a vortex due to gradients of chemical potentials ${\pmb \nabla} \mu_i$,
differs from zero.
This interference of diffusion and mutual friction has the following physical meaning:
diffusion affects particle velocities ${\pmb V}_i$
which, in turn, affect the vortex motion via the mutual friction mechanism (and vice versa).
\item
The dissipative cross-coefficient $\mathcal{B}_{pi}^\perp$, generally, differs from zero,
but vanishes in the first order in
$J_{e\mu} / \left| N_{{\rm V}p} D'_i \right| = (c J_{e\mu}) / (e_p n_i B)$
and
$D_i / \left| D'_i \right| = D_i / ( \pi \hbar n_i)$,
and thus can be neglected.
\item
The expression for $\mathcal{D}_{e\mu}^\parallel$, which describes diffusion of electrons and muons along the  vortex lines,
has exactly the same form as 
in the nonsuperfluid matter (see DGS20),
since the only force acting along the vortex line is the electron-muon friction.
\item
In contrast, the dominant first term in $\mathcal{D}_{e\mu}^\perp$ depends on the mutual friction parameters
$D_e$ and $D_\mu$.
This means that, for electrons and muons moving across the vortex array,
the momentum exchange between particles is mediated mainly by 
vortices [via the friction force, see the first term in Eq.~\eqref{eq:mf:Fj}],
instead of direct electron-muon interaction
[the term $J_{e\mu} \left( {\pmb V}_e - {\pmb V}_\mu \right)$ in the Euler equation \eqref{eq:mf:euler-e-mu}].
\item
The (nondissipative) coefficient
$\mathcal{D}_{e\mu}^H$ has, in the leading order, the same form as for nonsuperfluid matter.
This is not surprising, since this coefficient describes the Lorentz force acting on electrons and muons.
\end{enumerate}

\vspace{0.2 cm}
\noindent
%
{\bf Remark 1.}

If we consider another limit and neglect the friction force between flux tubes and electrons or muons,
i.e., set $D_e = D_\mu = 0$ (without assuming that $J_{e\mu}$ is small),
then diffusion and mutual friction are completely decoupled,
$B_{pi}^{\mu\nu} = C_{pi}^{\mu\nu} = 0$.
In addition, $\mathcal{A}_{pp}^\parallel$ and $\mathcal{A}_{pp}^\perp$ also vanish,
$\mathcal{A}_{pp}^\parallel = \mathcal{A}_{pp}^\perp = 0$,
so that the force on a vortex is described only by nondissipative coefficient
$\mathcal{A}_{pp}^H = - e_p n_p B / \left( c^2 T \right)$.
In turn, the generalized diffusion coefficients $D_{ik}^\parallel$, $D_{ik}^\perp$, and $D_{ik}^H$
in this approximation
take exactly the same form as in the nonsuperfluid matter (see DGS20).

In conclusion, we note that the presented scheme for calculating the phenomenological transport coefficients
can readily be generalized to arbitrary temperatures and particle compositions.

\bibliography{litt}

\end{document}